\documentclass{ametsoc}

\usepackage{float}
\usepackage{subfig} 
\usepackage{booktabs}
\newcommand{\comment}[1]{}
\allowdisplaybreaks
\makeatletter
\newcommand*\bigcdot{\mathpalette\bigcdot@{.6}}
\newcommand*\bigcdot@[2]{\mathbin{\vcenter{\hbox{\scalebox{#2}{$\m@th#1\bullet$}}}}}

\journal{mwr}

\bibpunct{(}{)}{;}{a}{}{,}

\title{Data-driven localization mappings in filtering the monsoon-Hadley multicloud convective flows} 

\authors{Mich\`ele De La Chevroti\`ere}
\affiliation{Department of Mathematics, the Pennsylvania State University, USA} 

\extraauthor{John Harlim \correspondingauthor{John Harlim, Department of Mathematics, The Pennsylvania State University, 109 McAllister Building, University Park, USA 16802.}}
\extraaffil{Department of Mathematics and Department of Meteorology, the Pennsylvania State University, USA}

\email{jharlim@psu.edu}

\abstract{This paper demonstrates the efficacy of data-driven localization mappings for assimilating satellite-like observations in a  dynamical system of intermediate complexity. In particular, a sparse network of synthetic brightness temperature measurements is simulated using an idealized radiative transfer model and assimilated to the monsoon-Hadley multicloud model, a nonlinear stochastic model containing several thousands of model coordinates. A serial ensemble Kalman filter is implemented in which the empirical correlation statistics are improved using localization maps obtained from a supervised learning algorithm.  The impact of the localization mappings is assessed in perfect model observing system simulation experiments (OSSEs) as well as in the presence of model errors resulting from the misspecification of key convective closure parameters. In perfect model OSSEs, the localization mappings that use adjacent correlations to improve the correlation estimated from small ensemble sizes produce robust accurate analysis estimates. In {\color{black} the presence of} model error, the filter skills of the localization maps trained on perfect and imperfect model data are comparable. } 

\begin{document}

\maketitle

%
\section{Introduction}

In high dimensional applications, ensemble Kalman filters are usually implemented using a small number of ensemble members due to the high cost in integrating the forecast model. This occurs for instance in operational numerical weather prediction, where forecast models have $O(10^8)$ state variables while computational resources can only allow to integrate up to $O(100)$ members \citep{houtekamerzhang2016}. With such small ensemble sizes relative to the state space dimension, the technical implementation of the EnKF can suffer from inaccurate state estimation, which manifests as underestimated forecast error covariances, spurious long-range correlations and ultimately filter divergence. The underestimation of forecast error covariances is often addressed by the method of covariance inflation \citep{anderson1999monte,anderson2007adaptive} while spurious correlations at long distances are usually mitigated using the technique of covariance localization \citep{hamill2001distance,houtekamer1998data}.  Localization schemes use tapering or kernel functions to reduce or zero out unphysical correlations, most often using a Schur product. The distance-based Gaspari-Cohn (GC) function \citep{gaspari1999construction} is a standard example of a parametric tapering achieved with a fifth order polynomial function on a compact support. The half-width of the compact support, which determines the geometric distance at which correlations are cut off, must be tuned for good performance.    

While parametric localization functions are useful in practice, they can be expensive to tune in large-scale applications. For instance, the GC localization function has an optimal half-width that tends to vary, e.g. by observation type \citep{houtekamer2005ensemble} and model variables \citep{anderson2007exploring,anderson2012localization} or even as a function of time \citep{anderson2012localization, chen2010cross}. Recently, \citet{de2017data} proposed a data-driven localization technique that {\color{black}can capture non-uniform localization bandwidths using a single parameter}. The technique uses ensemble archived products from which time series of sampled and undersampled correlations are computed. A supervised learning algorithm analyzes the two training correlation datasets to infer a localization function, named localization map. The localization map is used in verification mode to transform the poorly estimated sample correlation into an improved correlation. In a series of observing system simulation experiments (OSSEs) using the 40-variable Lorenz-96 model \citep{lorenz1996predictability} and a range of linear and nonlinear observation models, the localization maps were found to improve the filter estimates, most notably in the case of nonlinear indirect observations \citep{de2017data}.   

In light of these promising results obtained using a low-order model, the performance of the localization maps is further explored in a data assimilation system of intermediate complexity. Here, the serial least squares ensemble Kalman filter (LS-EnKF) of \citet{anderson:03} is implemented in the monsoon-Hadley multicloud model \citep{de2016zonally,mdlc2015}, a zonally symmetric model for the meridional Hadley circulation and monsoonal flow. The model's free troposphere synoptic scale wave dynamics is given by nonlinear equations for the barotropic and first two baroclinic modes of vertical structure, while the physical processes of convection and precipitation are represented by a stochastic model for clouds. Although the monsoon-Hadley multicloud model is an idealized atmospheric circulation model, it features a nonlinear multiscale wave dynamics with several thousand degrees of model coordinates, which makes it an ideal testbed for the localization maps. The vertical basis function representation of the model is exploited to recreate satellite-like observations using an idealized radiative transfer model. Brightness temperature-like measurements of six satellite channels are assimilated to the model using a sparse observational network. The filter skill of the localization maps is tested with this nonlinear indirect observation model in a series of perfect model OSSEs as well as in the presence of model error. 

The structure of this paper is as follows: In Section \ref{MHMM}, we review the general framework of the monsoon-Hadley multicloud model and look at numerical simulations of the model in two different regimes. In Section \ref{DAdesign}, the technique of the localization mapping is explained in the context of the LS-EnKF, followed by a description of the idealized radiative transfer model and general experimental design. In Section \ref{NumResults} we present the results of OSSEs realized in the perfect and imperfect model scenarios. We wrap up the paper with a brief summary and conclusions in Section \ref{SumCon}.

\section{The monsoon-Hadley multicloud model}\label{MHMM}
The monsoon-Hadley multicloud model is a zonally symmetric model for the large-scale Hadley circulation, ambient winds, and precipitation associated with the Summer monsoon season \citep{de2016zonally,  mdlc2015}.  The model is based on the Galerkin projection of the primitive equations of atmospheric synoptic dynamics onto the first modes of vertical structure in the free troposphere, and is coupled to a bulk atmospheric boundary layer (ABL) model. The prognostic variables of this vertical projection are the barotropic and baroclinic horizontal velocities, $\boldsymbol{u}_0, \boldsymbol{u}_1,  \boldsymbol{u}_2$, where each of these modes have zonal and meridional components, $u$ and $v$, respectively, and the baroclinic potential temperatures, $\theta_1,\theta_2$. The corresponding vertical basis, depicted as functions of height between the sea level height 0 to the tropopause $H_T=16 $ km, are shown in Figure~\ref{fig_vertical_struc}(a). The free tropospheric pressure $p$ and vertical velocity $w$ are given by Galerkin expansions consistent with the basis functions, with baroclinic wave mode amplitudes calculated diagnostically via the hydrostatic equation and incompressibility condition, respectively. Below the free troposphere model we place a mixed representation of the atmospheric boundary layer (ABL) with prognostic variables that include the horizontal velocity $\boldsymbol{u}_b =  (u_b, v_b)$, fluctuation potential temperature $\theta_b$ and equivalent potential temperature $\theta_{eb}$. See \citet{mdlc2015}, \citet{de2016zonally} and \citet{waite2009boundary} for the detailed formulation. The moist dynamics of the model is modeled with bulk equations for the vertically averaged water vapor free tropospheric moisture fluctuation $q$.

Altogether, the governing equations for the large-scale variables $\{ \theta_{eb}, \theta_b, \boldsymbol{u}_b, \boldsymbol{u}_0, \boldsymbol{u}_1, \boldsymbol{u}_2, \theta_1, \theta_2, q \}$ form the dynamical core of the monsoon-Hadley multicloud model. The dynamical core is a nonlinear  system of 13 partial differential equations (PDE) that is not conservative and not necessarily hyperbolic. The PDE system is solved iteratively with the Poisson equation for the ABL pressure $p_b$ using an operator time-splitting strategy \citep{strang1968construction}. The details of the numerical scheme is presented in \citet{de2016zonally}. The unresolved features of convection and precipitation, which will be described in Section~2\ref{smcm} below, are represented by the stochastic multicloud cumulus parameterization scheme of Khouider et al. \citep{khouider2010stochastic, khouider2006model, khouider2006multicloud, khouider2006simple, khouider2008multicloud,khouider2007simple}. 

The computational domain is reduced to a meridional slice of the troposphere between $40^{\circ}$S and $40^{\circ}$N (roughly 9000 km) with a mesoscale grid resolution of about 37 km. The system is integrated as an initial value problem with the  initial condition set to a radiative convective equilibrium (RCE;  \citet{mdlc2015}). This is a spatially homogeneous steady-state solution where the convective heating is balanced by the radiative cooling. Details on how to construct a RCE solution for the coupled system can be found in \citet{mdlc2015}. 
In Section 2\ref{monsoon} we present numerical simulations in an idealized boreal summer setting. A detailed description of the model can be found in the original works \citep{de2016zonally, mdlc2015}.
 
\subsection{The stochastic multicloud parameterization}\label{smcm}

The multicloud model highlights the role of three heating rates, $H_c$, $H_d$, and $H_s$, corresponding to the three cloud types that are observed to characterize organized tropical convection: Cumulus \textbf{c}ongestus cloud decks that heat the lower troposphere and cool the upper troposphere with cloud top near the freezing level, \textbf{d}eep convective towers that heat the entire troposphere, and \textbf{s}tratiform anvils that warm and dry the upper troposphere and cool and moisten the lower troposphere. These three cloud types are believed to be responsible for the bulk of the tropical rainfall and constitute a major source of heat for the free tropospheric circulation \citep{johnson1999trimodal,mapes2006mesoscale,abhik2013possible}. The convective heating rates $H_c$, $H_d$ and $H_s$ directly force the first two baroclinic modes of vertical structure as illustrated in Figure~\ref{fig_vertical_struc}(b). 

The parameterization scheme overlays on top of each grid box a Markov chain square lattice of size $n\times n$, where each lattice site is either cloud free or occupied by a congestus, deep or stratiform cloud, denoted as state 0, 1, 2 and 3, respectively. A continuous-time Markov process is then defined for each lattice element, allowing transitions from one state to another according to probability transition rates $r_{kl}$, $k,l=0,1,2,3$, that are Arrhenius-type functions of the convective available potential energy integrated over the whole troposphere (CAPE), low-level CAPE (CAPE$_l$), and midtroposphere dryness $D$. The probability rates $r_{kl}$ are constrained by a set of intuitive rules which are based on observations of cloud dynamics in the tropics \citep{johnson1999trimodal, mapes2006mesoscale}. For example, a clear sky site turns into a congestus site  with high probability if CAPE$_l$ is positive and the middle troposphere is dry, while a congestus site (or clear site) turns into a deep convective site with high probability if CAPE is positive and the middle troposphere is moist. The three cloud types are assumed to decay naturally into a cloud free site at some fixed rate. These rules are formalized in Table \ref{table0}. Note that since cloud transitions occur arguably on different time scales, the probability transition rates $r_{kl}$ are divided by their characteristic timescales $\tau_{kl}$. We use timescale estimate values resulting from a statistical Bayesian inference study of large-eddy simulated data \citep[see Table 3]{de2016stochasticity}. 

Assuming all $n^2$ Markov chains are independent, one can formally derive \citep{katsoulakis2003coarse, khouider2010stochastic,khouider2003coarse} the stochastic dynamics for the grid box cloud area fractions alone. Effectively, given the prescribed timescales $\tau_{kl}$ and the large-scale thermodynamic state (e.g. CAPE, $D$) at a grid box, the scheme outputs the {\color{black}time dependent} stochastic cloud area fractions $\sigma_c$, $\sigma_d$ and $\sigma_s$ for that column. These in turn influence the large-scale heating rates according to the following convective closure equations \citep{khouider2010stochastic, khouider2006simple, khouider2008multicloud}:
\begin{subequations}\label{closure_eqn}
\begin{align}
H_d =& \left\{\sigma_d \overline{ Q} + \frac{\sigma_d}{\overline{ \sigma_d }\tau^o_c} \left[a_1\theta'_{eb}+a_2q' -a_0 (\theta'_1+a_3 \theta'_2)\right]\right\}^+, \\
H_s =& \frac{\sigma_s\alpha_s}{H_m} \sqrt{ \mathrm{CAPE}_{}^+}, \\
H_c=&\frac{\sigma_c\alpha_c}{H_m} \sqrt{ \mathrm{CAPE}_l^+ }, 
\end{align}
\end{subequations}
where $x^+ = \max(x,0)$, $a_0, \dots, a_3$ and $Q$ are model constants and $H_m$ is the average middle troposphere height; $\tau^o_c$ is a reference convective timescale and $\alpha_s$, $\alpha_c$ determine the contributions of CAPE and CAPE$_l$ to congestus and stratiform heatings, respectively. The constant and parameter values in (\ref{closure_eqn}) can be found in \citet{de2016zonally}. The closure equations are expressed in terms of RCE quantities, which are denoted by overbars, and deviations from the RCE, denoted by primes (e.g. $\overline{Q}$ denotes the heating potential at RCE). 

\subsection{Idealized boreal summer monsoon simulations}\label{monsoon}
The monsoon-Hadley multicloud model is tested in an idealized summer monsoon setting on an aquaplanet with constant but nonuniform sea surface temperature (SST) mimicking the Indian and Pacific Oceans' warm pool (WP).  The prescribed SST follows a Gaussian meridional profile centered at $15^\circ$ N, as shown in Figure \ref{temp_profile}. This is meant to replicate the warm SSTs observed in the Intertropical Convergence Zone (ITCZ) during the boreal summer. We use a multicloud stochastic lattice of size $n=30$ embedded in our 37 km resolution meridional grid, which results in convective cells with a horizontal extent of $O$(1 km). 

The model is integrated for roughly 1350 days with a time step of 3 minutes. The solutions are in a statistical steady state after a short transient period of 100 to 200 days \citep{mdlc2015}. The first 1000 days are discarded as burn-in and the last 350 days are used for training purposes. Here we present the simulation results for two parameter regimes A and B, which differ only by their convective parameterization cloud timescales: regime A uses the Bayes mean timescale estimates of Table \ref{table0} while the timescales of regime B are obtained by adding one Bayes standard deviation to the mean. {\color{black}We should point out that most of the model errors is due to misspecification in the deep clouds decaying rate with large standard deviation.} We will use these two regimes to simulate numerical experiments with model error.

The mean meridional circulation resulting from taking the time average of the solution over the training interval is plotted in Figure~\ref{meancirc} for regime A. The {\color{black}height-latitude}  contour plots are shown for the horizontal wind components $u$ and $v$, vertical velocity $w$, potential temperature $\theta$, pressure $p$ and total heating $H$, each obtained from its respective Galerkin expansion as detailed in \citet{de2016zonally}. The cross sections show the dominant deep tropospheric overturning of the Hadley circulation, with an ascending branch over the WP at $15^\circ$N resulting from low-level convergence, and subsidence near $10^\circ$ S. The upward motion branch of the Hadley cell is associated with a strong deep barotropic heating mode and a stratiform second baroclinic potential temperature mode. The sea level pressure drops significantly moving northward through the ITCZ, a characteristic of the monsoon trough. The low-level wind displays the turning of the equatorial easterlies to westerlies South of the pressure trough and then back to easterlies, similar to the mean monsoonal flow of the boreal summer season. 

The Hovm\"oller plot diagrams of the wave fluctuations from the mean solutions are shown for regime A in Figure~\ref{waves}. The latitude-time contours of $\theta_{eb}$, $q$ and $v_0$ are plotted for a 25 day period starting at day 1200. The contour plots of the three cloud area fractions are also pictured during that same period. Small scale intermittent events can be observed throughout the domain, with a larger concentration over the WP,  roughly the $5-25^\circ$N band. Mesoscale cloud clusters are seen propagating southward and northward from the WP with suppressed and active phases of convection alternating every 1 to 2 days. We also observe that the strong events in the precipitation related fields $\theta_{eb}$ and $q$ correlate well with the cloud coverage peaks.  

The mean meridional circulation and Hovm\"oller plots for regime B are shown in Figures~\ref{meancirc2} and \ref{waves2}, respectively. As mentioned before, the cloud transition timescales of regime B are larger than those of regime A by one standard deviation. This positive bias in the transition timescales has an impact on the wave disturbances: cloud systems are now larger and persist over several days, while their period of oscillation is in the order of 10 days or so. The mean meridional circulation shows a reduced low level convergence and dampen upward motion  over the WP region. The convective total heating also appears diminished throughout the domain. 

\section{Data assimilation methodology and experimental design}\label{DAdesign}

The nonlinear Gaussian discrete-time filtering problem that we consider in this paper can be written in a compact form as follows,
\begin{subequations}
\begin{eqnarray}
\boldsymbol{x}_{m+1}&=& F(\boldsymbol{x}_m)\label{dynamics}\\
\boldsymbol{y}^o_{m}&=&G(\boldsymbol{x}_m)+ \boldsymbol{\epsilon}_{m}, \quad   \boldsymbol{\epsilon}_{m} \sim \mathcal{N}(0,R),\label{observation}
\end{eqnarray}
\end{subequations}
where $F$ is a nonlinear forecast model operator (representing here the Hadley monsoon stochastic multicloud model) and $G$ is a nonlinear observation model with a Gaussian measurement error $\boldsymbol{\epsilon}$ with zero mean and covariance $R$. The ensemble Kalman filter (EnKF) is an approximate filtering method introduced by \citet{evensen1994sequential} to estimate the first two-order statistics of the nonlinear filtering problem in \eqref{dynamics}-\eqref{observation}. The key idea of the EnKF is to model the background (or forecast) density as a Gaussian distribution with mean and covariance estimated empirically from an ensemble of $K$ forecast solutions. Subsequently, analysis mean and covariance matrix are computed using the Kalman filter formula \citep{kalman1960new}, incorporating the information from the current observations to the mean update. Finally, the analysis (or posterior) ensemble estimates are drawn from the Gaussian analysis distribution. We should mention that there are many variations of EnKF (e.g. \citet{houtekamer1998data,bishop:01,anderson:01,whitaker:02}) since there are non-unique ways to sample the Gaussian analysis  density.  

In theory, for {\color{black} linear models with Gaussian errors}, the EnKF converges to the (exact) KF solution in the limit of large ensemble size, $K \rightarrow \infty$. However the use of large ensembles is often cost prohibitive in practice, and typically we resort to small ensembles of size $K \ll N$, where $N$ is the dimension of the state space. For instance, in our application $K=O(10)$ while $N=O(10^3)$. While a small ensemble size is computationally desirable, it introduces sampling errors that lead to underestimated covariances \citep{furrer2007estimation}. This issue is routinely addressed  using \textit{covariance inflation} methods \citep{anderson2007adaptive} that essentially ``blow up'' the prior covariance by some inflation factor. Furthermore, small ensemble sizes also induce spurious ensemble correlations between observations and state variables with distant grid points \citep{lorenc2003potential}. This problem is usually mitigated using a technique known as \textit{localization} \citep{houtekamer1998data}. 

In this paper, we address the spurious correlation issue with the data-driven technique proposed by \citet{de2017data}. Instead of tuning a specified parametric localization function (such as the half-width parameter of the Gaspari-Cohn or exponentially decaying functions), this method uses a pair of labeled training datasets to estimate a linear map, called the localization map, that transforms the poorly estimated sample correlation into an improved correlation. This training methodology is effectively an example of supervised learning in the machine learning community \citep[see e.g.][]{friedman2001elements}. Here, the localization map is implemented within the sequential least squares framework of Anderson (LS-EnKF; \citet{anderson:03}), a serial variant of the EnKF which allows for observations with independent measurement errors to be assimilated sequentially. Anderson's scheme breaks down the filtering problem into a sequence of linear regressions of a scalar observation onto the state vector. In this scalar context, the covariances, or correlations, between a single observation and the model state variables appear explicitly and can be easily localized. 
 
In the remainder of this section, we provide a brief review on the LS-EnKF in Section 3\ref{lsenkf} and the localization map technique in Section 3\ref{locmap}. In Section 3\ref{RTM}, we introduce the idealized radiative transfer model for synthetic satellite observations and conclude with the experimental design in Section 3\ref{designosse}. 

\subsection{Least squares EnKF algorithm}\label{lsenkf}
In the LS-EnKF, an ensemble of $K$ model state samples is integrated forward using the forecast model $F$, producing a set of prior (or forecast) solutions $\{\boldsymbol{x}^{f,k}  \}_{k=1}^K$ (note that the time index is suppressed for clarity of notation). Each ensemble member $\boldsymbol{x}^{f,k}$ is then projected to the observation space by applying the observation operator, i.e. $\boldsymbol{y}^{f,k} = G(\boldsymbol{x}^{f,k})$. The  mean and variance of the $j^{\mathrm{th}}$ component of the observation vector $\boldsymbol{y}^{f,k}\in \mathbb{R}^M$ are approximated by its ensemble mean and variance,
\begin{equation}
 \overline{y}^f_j =\frac{1}{K} \sum_{k=1}^K y^{f,k}_j,\quad\quad\mbox{and }\label{sample3}
\qquad P^f_{y_jy_j}=\frac{1}{K-1} \sum_{k=1}^K (y^{f,k}_j -  \overline{y}^f_j)^2, \qquad j=1,\dots,M,
\end{equation}
respectively. The LS-EnKF algorithm works under the assumption that observation errors are independent (the matrix $R \in \mathbb{R}^{M\times M}$ is diagonal) and uses the unperturbed observation component $y^o_j$ to sequentially update the ensemble solutions. {\color{black}We should point out that our choice of using this algorithm is just for convenience, one can consider more advanced data assimilation techniques that can handle correlated observation errors, especially when dealing with satellite measurements \citep{waller2014}.} For each observation component $j$, $j=1,\dots,M$, the LS-EnKF executes the following update:
\begin{subequations}\label{updatey}
\begin{eqnarray}
 \overline{y}^a_j &=& \overline{y}^f_j + (P^f_{y_jy_j}+R_{jj})^{-1}P^f_{y_jy_j} (y^o_j-\overline{y}^f_j),\\
y^{a,k}_j &=& \overline{y}^a_j + \sqrt{\frac{R_{jj}}{R_{jj}+P^f_{y_jy_j}}} (y^{f,k}_j-\overline{y}^f_{j}),
\end{eqnarray}
where $R_{jj}$ is the $j^{\mathrm{th}}$ element of the diagonal of $R$. Note that the update step (\ref{updatey}) can be interpreted as a scalar ensemble filter in the observation space. The second step of the LS-EnKF is to regress the increment $\Delta y^k_j = y^{a,k}_j -y^{f,k}_j $ of the observation variable onto the state variables as follows:
\begin{eqnarray}
\boldsymbol{x}^{a,k} &=& \boldsymbol{x}^{f,k} + \frac{P^{f}_{\boldsymbol{x}y_j}}{P^f_{y_jy_j}}\Delta y^k_j,\label{lls} 
\end{eqnarray}
\end{subequations}
where the cross-covariance vector $P^{f}_{\boldsymbol{x}y_j} \in \mathbb{R}^N$ is also approximated by its ensemble statistics, that is,
\begin{equation}\label{sample2}
P^{f}_{x_iy_j} =\frac{1}{K-1} \sum_{k=1}^K (x^{f,k}_i -  \overline{x}^f_i)(y^{f,k}_j -  \overline{y}^f_j), \qquad i=1,\dots,N.
\end{equation}
As mentioned before, the use of small ensemble sizes $K \ll N$ has two adverse effects: 1) underestimating the covariance and 2) producing unphysical, spurious long-range correlations. In our numerical implementation, the first issue is dealt with using the adaptive covariance inflation of \citet{anderson2007adaptive} while the second is addressed using the localization mapping technique described in the next section. 

\subsection{Localization mappings and modified LS-EnKF} \label{locmap}
The localization mappings introduced in \citet{de2017data} use information from the sample correlation matrix between the model variable $\boldsymbol{x}$ and the observation variable $\boldsymbol{y}$, defined in the usual way as
\begin{equation}\label{corr}
C_{\boldsymbol{x}\boldsymbol{y}}^K = \mathrm{diag}(P_{\boldsymbol{x} \boldsymbol{x}})^{-1/2} P_{\boldsymbol{x}\boldsymbol{y}}\, \mathrm{diag}(P_{\boldsymbol{y} \boldsymbol{y}})^{-1/2} \,\, \in \mathbb{R}^{N\times M}. 
\end{equation}
Here, $\mathrm{diag}(P_{\boldsymbol{x} \boldsymbol{x}}) $ and $\mathrm{diag}(P_{\boldsymbol{y} \boldsymbol{y}})$ are diagonal square matrices having as their main diagonal the diagonal elements of $P_{\boldsymbol{x} \boldsymbol{x}}$ and $P_{\boldsymbol{y} \boldsymbol{y}}$, respectively. For example, the diagonal elements of $\mathrm{diag}(P_{\boldsymbol{y} \boldsymbol{y}})$ are given as in  (\ref{sample3}). The elements of the cross-covariance matrix $P_{\boldsymbol{x}\boldsymbol{y}}$, on the other hand, are given by (\ref{sample2}). All are estimated from an ensemble of size $K$. 

The idea behind the localization mappings is to obtain a map $\mathcal{L}$ that transforms, at each time step $m$, a poorly estimated correlation matrix $C_{\boldsymbol{x}\boldsymbol{y}}^{K,m} $ obtained using a small ensemble of size $K\ll N$ into an improved correlation matrix, that is ``closer'' to some target  correlation $C_{\boldsymbol{x}\boldsymbol{y}}^{L,m}$ estimated using a large ensemble size $L \gg K$. In practice, $L$ is taken as large as possible, in such a way that $C_{\boldsymbol{x}\boldsymbol{y}}^{L,m}$ is a good approximation to the asymptotic correlation $C_{\boldsymbol{x}\boldsymbol{y}}^{\infty,m}$, at time $m$. 

The map $\mathcal{L}$ is assumed to be of linear form, and we seek an estimate of $\mathcal{L}$ that minimizes the expected squared error between the transformed correlation and the target correlation $C_{\boldsymbol{x}\boldsymbol{y}}^{L}$. Specifically, for every pair of observation $j$ and model state variable $i$, we find $\mathcal{L}_{ij} \in \mathbb{R}^{N_{loc}}$, $N_{loc} < N$, that minimizes the following error cost function
\begin{equation}\label{minprob1}
J(\mathcal{L}_{ij}) = \int_{[-1,1]^{2\rho + 2}} \left[ \sum_{\ell= i - \rho}^{i + \rho}  \mathcal{L}^{\ell}_{ij} C_{\ell j}^K - C^L_{ij} \right]^2 p( C_{\bigcdot j}^K \big| C_{ij}^L) p(C_{ij}^L) \, dC_{\bigcdot j}^K    dC_{ij}^L,  \qquad \mathcal{L}_{ij} \in \mathbb{R}^{N_{loc}},
\end{equation}
where $C_{ij}:=C_{x_i y_j}$, $\mathcal{L}_{ij}^\ell$ is the $\ell^{\,\mathrm{th}}$ component of $\mathcal{L}_{ij}$, and $C_{\bigcdot j} := (C_{(i-\rho)j}, \dots, C_{(i+\rho)j}) \in \mathbb{R}^{N_{loc}}$. {\color{black} Here, we emphasize $J$ as a function of $\mathcal{L}_{ij}$ although it also depends on $N_{loc}$ (or $\rho$). }The distributions of the densities $p(C^L_{ij})$ and $ p( C^K_{\bigcdot j} \big| C^L_{ij})$ are assumed to be stationary and will be sampled from a training dataset. Effectively, the solution of the minimization problem (\ref{minprob1}) produces an estimate of the correlation $C^L_{ij}$ through a linear map of size $N_{loc}=2\rho + 1$ that combines the information of the local subsampled correlations $C^K_{\ell j}$ which are spatially located within a radius $\rho$ of the model variable $x_i$'s grid location. A special form of (\ref{minprob1}) arises when the radius $\rho=0$ and $\mathcal{L}_{ij} \in \mathbb{R}$ is a scalar learned from the point correlation $C_{ij}^K$. In this case,  the data-driven map $\mathcal{L}  = \{ \mathcal{L}_{ij}\}_{i,j=1}^{N,M }$ corrects the correlation matrix $C_{\boldsymbol{xy}}$ pointwise,  in the same manner as a Schur product.

One way to approximate the solution of the minimization problem in (\ref{minprob1}) is to discretize the cost function using samples from $p( C_{\bigcdot j}^K \big| C_{ij}^L)  $ and $p(C^L_{ij})$. We next describe a sampling method based on a historical data assimilation product using a large ensemble size $L\gg K$. {\color{black} In high-dimensional applications, one may not have this training data set since it is computationally not feasible to obtain a reasonably accurate state estimation by employing EnKF with a very large $L$ without any form of localization.} In this situation, then one can simulate the training data using an EnKF with an ensemble size $L$ that is only slightly larger than the verification size $K$ (e.g. $L=\mathcal{O}(10^2-10^3)$ for $K=10$), with {\color{black} a very broad localization range} to obtain a reasonably accurate state estimation. {\color{black}An example of such experiment was reported by \citet{miyoshi2014} with 10,240 ensemble members.}

Our sampling method first consists of generating training data using a global EnKF scheme with an ensemble of size $L$. Suppose that for each ensemble member $k=1,\ldots,L$, the assimilation experiment generates a time series of forecasts $\{ \boldsymbol{x}_1^{f,k}, \boldsymbol{x}_2^{f,k}, \dots, \boldsymbol{x}_T^{f,k}  \}$, where $T$ is the number of assimilation cycles.  Then, at each cycle $m$, $m=1,\dots,T$, we calculate the sample correlation $C^{L}_{ij,m}$ using all of the ensemble members as well as a \textit{subsampled} correlation $C^{K}_{ij,m}$, selecting only $K$ members out of $L$. By this method, we effectively obtain samples $C^{L}_{ij,m} \sim p(C_{ij}^L)$ and $ C^K_{\bigcdot j, m} \sim p( C^K_{\bigcdot j} \big| C^L_{ij})$. Given these samples, the Monte-Carlo approximation to the minimization of the integral equation in (\ref{minprob1}) is given by 
\begin{equation}\label{minprobdiscrete}
\min_{\mathcal{L}_{ij}} \frac{1}{T} \sum_{m=1}^T \left[ \sum_{\ell=i - \rho}^{i + \rho} \mathcal{L}_{ij}^{\ell} C^K_{\ell j,m} - C^L_{ij,m} \right]^2, \qquad \mathcal{L}_{ij} \in \mathbb{R}^{N_{loc}},
\end{equation}
which is essentially a linear least squares problem. {\color{black}For every $i, j$, the explicit solution of this linear least-square problem is given by ${\boldsymbol{x}}=(A^\top A)^{-1}(A^\top {\bf b})$, for ${\boldsymbol{x}} = ({\mathcal{L}^\ell_{ij}})\in\mathbb{R}^{2\rho+1}$, $A=(C^{K}_{\ell j,m})\in\mathbb{R}^{T\times (2\rho+1)}$ and ${\bf b} = (C^L_{ij,m})\in\mathbb{R}^T$.} Solving (\ref{minprobdiscrete}) for all $i$ and $j$, we obtain the linear map estimator $\hat{\mathcal{L}}=\{ \hat{\mathcal{L}}_{ij}\}_{i,j=1}^{N,M}$. Finally, the straightforward modification on the LS-EnKF is to replace $P^f_{\boldsymbol{x}y_j} $  in Eqn. (\ref{lls}) by 
\begin{equation}\label{mod}
P^f_{\boldsymbol{x}y_j} =  \sqrt{P^f_{y_j y_j}} \mathrm{diag}(P^f_{\boldsymbol{x} \boldsymbol{x}})^{1/2}  C_{\boldsymbol{x}y_j}^K \leftarrow \sqrt{P^f_{y_j y_j}} \mathrm{diag}(P^f_{\boldsymbol{x} \boldsymbol{x}})^{1/2}   \hat{C}^L_{\boldsymbol{x} y_j}.
\end{equation}
Here, the components of $ \hat{C}^L_{\boldsymbol{x} y_j}$ are given by $ \hat{C}^L_{x_i y_j} = \sum_{\ell= i - \rho}^{i + \rho}  \hat{\mathcal{L}}^{\ell}_{ij} C_{\ell j}^K $. The new and improved correlation $ \hat{C}^L_{\boldsymbol{x} y_j}$ is closer, in the least square sense, to the target correlation $C_{\boldsymbol{x}y_j}^L$. Note that the map $\mathcal{L}$ is trained offline on the data $\{C^{K}_{\boldsymbol{x}\boldsymbol{y},m}, C^{L}_{\boldsymbol{x}\boldsymbol{y},m} \}_{m=1}^T$ and the computational complexity of each regression problem is $\mathcal{O}(LN_{loc}T)$ \citep{de2017data}. {\color{black} The training of $\mathcal{L}$ with $0 \le \rho \le 6$ using a Matlab serial application takes an approximate 13 hours wall-clock time on a single CPU. The relative residual norm of the linear estimator $({\hat{\mathcal{L}}^\ell_{ij}})= \hat{\boldsymbol{x}} $, calculated as $\| A \hat{\boldsymbol{x}} -  \boldsymbol{b} \|_{2} / \| \boldsymbol{b} \|_{2}$, is on average 20\% and is weakly monotonic decreasing as a function of $\rho$ or $K$.} 

\subsection{An idealized radiative transfer model}\label{RTM}
In our numerical experiments below, we consider assimilating satellite-like observations based on an idealized radiative transfer model that assumes an \textit{absorption coefficient} of the form
\begin{equation}
\alpha(q,z) = \alpha_0 \exp[- z /(H_T\,  \tilde{q} )], \label{AC}
\end{equation}
where $\alpha_0$ is a reference value, $H_T$ is the height of the troposphere and $\tilde{q}$ is a rescaled measure of the free troposphere vertically averaged moisture anomaly $q$ (in the experiments reported here, $\tilde{q} = 0.1( q - a )/(b-a) + 0.05$, where $a = \min(q)$ and $b = \max(q)$ and {\color{black} the extrema are taken over the training data set}).  As expected, $\alpha$ decreases exponentially with height, and is sensitive to the atmospheric moisture content $q$. The \textit{optical thickness} of the atmosphere between heights $z$ and $z_1$ is defined as
\begin{equation}
\tau(q,z_1,z) = \int_z^{z_1} \alpha(q,z') dz' {\color{black} = H_T \tilde{q}[\alpha(q,z) - \alpha(q,z_1)]} , \label{OT}
\end{equation}
where the last equality is obtained via the integration of the absorption coefficient (\ref{AC}). The  \textit{transmittance} $T_\lambda$ at  the wavelength $\lambda$ is given by a semblance of \textit{Beer's Law}:
\begin{equation}
T_\lambda(q,z_1,z) = \exp (-\tau(q,z_1,z)) = \exp( H_T \, \tilde{q} [ \alpha(q,z_1)-\alpha(q,z)]).
\end{equation} 
The vertical derivative of the transmittance is known as the \textit{weighting function} $K_{\lambda}$: 
\begin{equation}
K_{\lambda}(q,z_1,z) = \frac{\partial T_\lambda (q,z_1,z)}{\partial z} = \alpha(q,z) T_\lambda(q,z_1,z).    \label{WF}
\end{equation}

We desire to simulate channels whose weighing function peaks at a specific height $z_{\max}$ in the troposphere. This occurs at  $z_{\mathrm{max}} = H_T \tilde{q}  \ln[\alpha_0  H_T \tilde{q} ]$. Using this critical height we determine $\alpha_0 =  \exp[z_{\mathrm{max}}/(H_T \tilde{q} ) ]/(H_T \tilde{q})$, which in turn specifies $K_{\lambda}$. We select 6 distinct wavelengths $\lambda_1, \dots, \lambda_6$, that we also call \textit{channels}, associated with the heights  $z_{max} = 2, 4,6,8,10,12$ km, respectively. Figure \ref{SixChannels} shows the top-of-atmosphere optical thickness, transmittance, and weighting function for these 6 channels for various atmospheric states drawn from climatology. 

The integrated brightness temperature at height $z_1$ associated with the wavelength $\lambda$ is modeled by 
\begin{eqnarray}
 T_{b,\lambda}(\theta,\theta_b,q,z_1) = \theta_b T_\lambda(q,z_1,0)+  \int_0^{z_1} \theta(z)  K_\lambda (q,z_1,z) \, dz \label{SW5}
\end{eqnarray}
where $\theta_b$ is the ABL potential temperature and $\theta(z)$ is the potential temperature at height $z$ reconstructed by linear superposition of its first two baroclinic modes as described in Figure~\ref{fig_vertical_struc}. 
For our data assimilation experiments, we consider the top-of-atmosphere satellite brightness temperature-like measurements defined as $T_{b,\lambda}(\theta,\theta_b,q,\infty)$. Figure \ref{BT} shows the top-of-atmosphere brightness temperature $T_{b,\lambda}(\theta,\theta_b,q,\infty)$  for the channel $\lambda_1$ calculated from the climatology over a period of 65 days. 

\subsection{Experimental design}\label{designosse} 
The monsoon-Hadley multicloud model described in Section \ref{MHMM} coupled with the modified LS-EnKF given by equations (\ref{updatey}) and (\ref{mod}) form our assimilation scheme. We will conduct OSSEs in the perfect as well as imperfect model scenarios. In the perfect model scenario, both the nature and forecast states are generated using the same model configuration (regime A; identical twin experiments). In the imperfect model scenario, the nature is generated using the reference parameters of regime A while the forecast model parameters are of regime B, as discussed in Section 2\ref{monsoon}.

The analysis is performed over the 256 internal grid points of the meridional numerical domain, on the 14 large-scale fields $\{ \theta_{eb}, \theta_b, p_b, \boldsymbol{u}_b, \boldsymbol{u}_0, \boldsymbol{u}_1, \boldsymbol{u}_2, \theta_1, \theta_2, q\}$. Thus the model state is of dimension $N=14\times 256=3584$. The stochastic cloud area fractions $\sigma_c$, $\sigma_d$ and $\sigma_s$ of the convective parameterization are not filtered given the added algorithmic complexity of constraining the updated cloud states to the nonnegative range. {\color{black}Interested readers should consult the alternative method proposed in \cite{janjic2014}, which was designed to preserve the positivity and conserve mass.} After analysis, the cloud heating rates $H_c$, $H_d$ and $H_s$ are calculated using the background cloud area fractions and the analyzed large-scale fields to enforce the convective closure balance of the multicloud parameterization. 

The observations are generated from the nature run according to \eqref{observation}, where the observation model $G$ is the idealized radiative transfer model described in the previous section, and the observation error covariance matrix $R$ is diagonal with components equal to $10\%$ of the climatological variances. More precisely, $G$ maps the mass field $\{\theta_{b},\theta, q\}$ at an observation location to a brightness temperature-like measurement $T_{b,\lambda_\ell}(\theta,\theta_b,q,\infty)$ for the channel $\lambda_\ell$, $\ell=1,\dots,6$. All 6 channels are observed on 64 uniformly distributed meridional locations (every 148 km or so) for a total of $M = 6 \times 64=384$ observations. Observations are assimilated at every 30 model integration steps (1.5 hr). 

The correlation data used to train the localization maps are obtained by running an OSSE using $L=1000$ members for about 1350 days, using the last 90 days (or $T=90\times 24 \div 1.5=1440$ cycles, accounting for an analysis every 1.5 hr) for training. At each cycle $m=1,\dots,1440$, we obtain correlation matrices $C^{1000}_m$ and $C^{K}_m$ for values of $K$ ranging from 10 to 35. The localization maps are obtained from solving the least square problem (\ref{minprobdiscrete}) using $\rho=6$. {\color{black} Results using other values of $\rho$ will be reported in some cases.} This means that we correct each correlation between a model state $x_i$ (say $u_0$ at the Equator) and an observation $y_j$ (radiance at some remote station) using a linear combination of the correlations between that observation and like-field model states ($u_0$) located within 6 grid points from the model state location (the Equator). {\color{black}In Figure~\ref{lmap_contour}, we show the vector localization maps for the correlations between the model coordinates $i$ of the field $u_b$ and a Channel 1 brightness temperature observation $j$ located on the green point (near the Equator). Each panel of this figure corresponds to the resulting map optimized for different $K$. Each vertical slice of the contour plot in each panel consists of 13 values ($\ell=i-\rho,\ldots,i+\rho, \rho=6$) obtained from solving the regression problem in \eqref{minprobdiscrete}. For {\color{black}the} observation at location $j$ (green dot), we solve the regression problems corresponding to the model grid points $i$ which satisfy $|i-j|\leq 32$, as shown in the horizontal axis in each panel of Figure~\ref{lmap_contour}. This choice is to avoid excessive computational storage and to ensure that we can capture the nontrivial nonlocal structure of the correlations. 

Notice that from Figure~\ref{lmap_contour} the structure of the map is not symmetric with respect to the observation location. Secondly, the function value of the map for larger $|i-j|$ (or the support) increases as a function of $K$. This is consistent with the results in a simpler context \citep{de2017data}. In fact, if $L=K$ in \eqref{minprobdiscrete}, the resulting map $\mathcal{L}^\ell_{ij}$ is one for $\ell=0$ and zero otherwise for all $i,j$. This means that for the case of $L=K$, the map does not provide any localization. If this map is used for filtering with $K<L$, then the filter will diverge. We will support this argument with numerical results below. 
}

{\color{black}Although the focus will be on the vector maps obtained with $\rho=6$, we will also show results of the scalar map ($\rho=0$) in the perfect model case as reference. In Fig.~\ref{GCvsLd}, we show examples of the scalar localization maps optimized for $K=15$ compared to the Gaspari-Cohn with half-width parameter equals to 6. Notice that the amplitudes and supports of the data-driven maps vary as functions of the observation location and variable. }

The accuracy of the assimilation is measured out of sampling (on a verification interval that is independent of the training data) by the time mean of the RMS error between the analysis ensemble mean, $\overline{\boldsymbol{x}}^{\,a}$, and the truth run, $\boldsymbol{x}^{\mathrm{truth}}$:
\begin{equation}
\mathrm{RMSE} = \left[\frac{1}{T_V} \sum_{m=1}^{T_V} (\overline{\boldsymbol{x}}_m^{\,a} - \boldsymbol{x}_m^{\mathrm{truth}})^2  \right]^{1/2},\nonumber
\end{equation}
where  $T_V$, the length of the verification interval, is set to 1 year or 365 days ($T_V=5840$ analysis cycles). 

\section{Numerical Results}\label{NumResults}

We now present the numerical results for OSSEs realized in the perfect and imperfect model scenarios. In the perfect model experiments, both the truth and forecast states are simulated with regime A. In the imperfect model experiments, the forecast model and the truth are simulated using regimes B and A, respectively (see Table~\ref{table1} for details).
Recall that regime B differs from regime A by its slightly larger convective parameterization timescales. The goal of the imperfect model experiments is to test the robustness of the localization maps in the presence of model error.

{\color{black} In the first numerical experiment, our goal is to check the sensitivity of the filter estimates on parameter $\rho$. To do this, we train the maps for $\rho=0-10$ for the case of $K=10$. In Figure~\ref{degradation}, we compute the average relative degradation as a function of $\rho$. Here, the average relative degradation is defined as follows,
\begin{equation}
RD_\rho = \frac{1}{14}\sum_{j=1}^{14} \frac{\mathrm{RMSE}_{j,\rho} - \min_\rho  \mathrm{RMSE}_{j,\rho}}{\min_\rho \mathrm{RMSE}_{j,\rho}},\nonumber
\end{equation}
where $RMSE_{j,\rho}$ denotes the RMSE for the $j$th variable (of the 14 components) and obtained using localization map with parameter $\rho$. The minimum is taken over the range of $\rho=0-10$. Based on this metric, one can see that the relative degradations for $\rho=5-10$ are on average 10\% less that those of $\rho=0-4$. Based on this empirical result, we will focus on the case of scalar map $\rho=0$ and vector map $\rho=6$ in the remaining of this paper.
}

\subsection{Perfect model experiments}

We first realize a perfect model experiment using 1000 members (TD\_PM), calculating the background correlation at each analysis cycle. Using this dataset we obtain two different maps: 1) A scalar map with $\rho=0$ and 2) a vector map with $\rho=6$. Examples of these two maps are shown in Figures~\ref{lmap_contour} and \ref{GCvsLd}. We summarize these (and the model error) experiment configurations in Table~\ref{table1}.
The results of the perfect model verification experiments using these two maps {\color{black}(labelled PM $\rho = 0$ and $\rho = 6$)} were shown in Figure \ref{perfectmodel}. We report the time mean analysis RMSE over a one year verification interval as a function of ensemble size for the 14 analysis fields.  For reference, a verification experiment using a GC localization with a half-length equal to 6 is included. Figure~\ref{perfectmodel} also reports the RMSE of the experiment TD\_PM and the climatological standard deviation which quantifies the error without data assimilation.  

{\color{black}Comparing the scalar map PM $\rho = 0$ with the vector map PM $\rho=6$ and the GC localization}, we find that the performance of {\color{black}the vector map PM $\rho=6$} is the closest to the training experiment TD\_PM realized with 1000 members. The improvement of {\color{black}the vector map} over {\color{black}the scalar map} is most markedly seen for the fields $u_0$ and $u_b$. In fact we observe that, of all fields, $u_0$ and $u_b$ have the highest RMSEs relative to climatology. Going back to the training data TD\_PM, an inspection of the autocorrelation function for $u_0$ and $u_b$ reveal that their signal is still strongly correlated over the 90 day training interval, which is a violation of the stationarity assumption of the cost function formulation in \eqref{minprob1}. {\color{black} We should also point out that the dynamics of these variables are almost constant (since they are slow) within the observation time scales (1.5 hours), which means that the linearized dynamics are close to identity (or marginally unstable with eigenvalues close to one). The basic Kalman filter theory suggests that the observability and controllability conditions are necessary for accurate estimation \citep{kalman1960new}. In our case, the observability condition is most likely violated since we don't observe $u_0$ and $u_b$ directly. The fact that the vector map PM $\rho = 6$ improves the estimates relative to the scalar map PM $\rho=0$ can be due to an improved controllability condition of the filter.  We should point out that an analogous finding (inaccurate estimate) was also reported by \cite{tardif2014} in the context of assimilating purely atmospheric data at frequent times in a coupled atmospheric-ocean data assimilation. To resolve this issue, they proposed to change the observation function and reduce the observation frequency, which effectively improve the observability condition (as an alternative to improving the controllabilty condition). While both filtering problems considered here and in \cite{tardif2014} are nonlinear, yet the classical linear filtering theory seems to give a plausible explanation for the results.} We also note that GC performs the worse among these numerical experiments. In fact, {\color{black}GC numerically blows up} when the ensemble size is too small (10 ensemble members in our experiment). 

\subsection{Model error experiments} We next investigate the performance of the localization maps in the presence of model error. We first run a model error experiment using 1000 members, producing the training dataset TD\_ME. {\color{black}We train two different maps on {\color{black}TD\_ME's} background correlations to be used in verification model error experiments (labeled ME2): 1) A scalar map ($\rho=0$) and 2) a vector map using $\rho=6$.} 

As a reference, we show the model error experiment ME1 using a ``perfectly tuned'' {\color{black}vector map ($\rho=6$)}, that is, a map trained on the perfect model training data TD\_PM. While this configuration is not practical in real applications since one does not have the knowledge of the true dynamical parameters (regime A in this example), this experiment provides, as we will discuss below,  an intuition of how sensitive the proposed method is to the model error in the training dataset.

The results of the model error verification experiments using the maps ME1 and ME2 as well as a GC localization (with a half-width equal to 6) are shown in Figure~\ref{modelerror}, along with the RMSE of the two reference training experiments TD\_PM and TD\_ME. The perfect model experiment {\color{black}PM $\rho=6$} is added for sake of comparison. {\color{black} Overall, GC performs the worst with numerical blow up at $K=10$. The scalar map ME2 $\rho=0$ produces improved filter estimates over GC except on $\theta_{eb}, \theta_1, q$ but it converges for the case of $K=10$. The vector map ME2 $\rho=6$ beats these two cases on all counts.} Also, the filtering skills of {\color{black}the vector} maps ME1 $\rho=6$ and ME2  $\rho=6$ are visually indistinguishable. For some of the fields, most notably for $v_b$, $v_0$, $v_2$ and $\theta_2$, their skills are close to that of the perfect model {\color{black}vector map} experiment {\color{black}PM $\rho=6$}. Interestingly, in some cases (e.g. $u_b, u_0, u_1, v_1, u_2$ and $\theta_1$) the vector maps ME2 outperform their own training experiment TD\_ME, especially when the ensemble size is large. {\color{black} This result reminisces the same finding by \cite{oke2007} in a simpler context with a perfect and linear model scenario. In particular, they found that localization on EnKF can improve the effective rank of the ensemble and outperform EnKF with large ensemble size without localization.}

{\color{black}While the results from using the {\color{black}vector maps ME2 $\rho=6$} are almost indistinguishable compared to ME1 {\color{black}$\rho=6$} in almost all cases,} we should note that the results of ME2 {\color{black}$\rho=6$} can be sensitive to the training data. In particular in the case $K=30$ we found that for the specific training data set, the resulting map produces filter divergence estimates. {\color{black} In this case, we found that the issue can be overcome using the map trained on a longer data set (6 months rather than 3 months; see the red markers in Figure~\ref{modelerror} for the case $K=30$). This sensitivity issue, which occurs in the experiments with larger $K$, can be understood as follows. {\color{black}First, the support of the resulting map grows as $K$ increases as shown in Figure~\ref{lmap_contour}. Second,} it is generally difficult to estimate correlations between two random variables that have small true correlations. Even in the simplest case, it is well known that the empirical correlation estimates of independent and identically distributed Gaussian variables with true correlation $\rho$ has error variances with leading order term $(1-\rho^2)^2K^{-1}$ \citep{hotelling:53}. This means that the estimates $C^K_{ij}$ for which the true correlations are small have large error variances. On the other hand, accurate estimates of the correlations, especially for those corresponding to $i$ and $j$ with large $|i-j|$ for maps with larger support, are crucial in the regression in \eqref{minprobdiscrete} for large $K$. Since the error of the Monte-Carlo approximation \eqref{minprobdiscrete} of \eqref{minprob1} is proportional to the square-root of the ratio between the variance of the integrand and the size of data $T$, then it is clear that larger $T$ is required to offset the larger variances. This is why the simulations with longer training dataset can overcome the issue. Third, the condition number of the least square problems in \eqref{minprobdiscrete} ranges betwen $10^1$ and $10^7$ (see e.g. Section 3.3 of \citet{demmel:97} for the definition of the condition numbers for linear least squares problems). This means that the proposed least square method is ill-conditioned and thus small perturbation (e.g., on the order of $10^{-7}$) to the data (due to model error) can yield an order one relative error in the resulting vector map. Indeed, when we compare the vector maps of the case $K=30$ that are trained on different training intervals (one that gives accurate analysis and another one that does not converge), the relative error of the resulting vector maps (in uniform norm) is on the order of one.

If longer data set is not available, we numerically found that one can also overcome this issue by choosing different $\rho$ (results are not shown). {\color{black}We should also mention that for a given $C^L_{ij}$, one can generate more samples of $p(C^{K}_{\bigcdot j}|C^L_{ij})$ in \eqref{minprob1}. Specifically, one can construct $C^K_{\bigcdot j}$ using different choices of $K$ ensemble members among the available $L$ training ensemble solutions. In this manuscript, we only regress to one sample of $p(C^{K}_{\bigcdot j}|C^L_{ij})$ for each $C^L_{ij}$. The point we want to make is that one can increase the training data by regressing each $C^L_{ij}$ to multiple $C^K_{\bigcdot j}$ constructed from different subsets of the training ensemble members.} 

Alternatively, one can use the maps optimized for smaller ensemble sizes.  As a supporting argument, we test the robustness of the localization {\color{black}vector} maps {\color{black}($\rho=6$)} by running a model error verification experiment with an ensemble size $K$, using a map ME2 optimized for an ensemble size $K' \ne K$. The RMSEs for $K, K'=10,15,\dots,35$ are reported in Figure \ref{swap}. The case $K=K'$ is identical to the experiment ME2 {\color{black}$\rho=6$} in Figure \ref{modelerror}. The results in Figure \ref{swap} reveal that filter fails systematically when $K'>K$, that is, when the training ensemble size is greater than the verification ensemble size. {\color{black}This result is consistent with our explanation above. That is, since the map's support increases as a function of $K'$, the failure} in the case of $K'>K$ is partially due to the existing spurious, long-range correlations that are not damped out by the map with larger support trained with ensemble size $K'$. On the other hand, except for $u_b$ and $u_0$, the filtered estimates monotonically degrade if one uses the maps trained on $K'<K$ to filter with an ensemble of size $K$. In this case, the maps trained with smaller ensemble size $K'<K$ overly damp out the spurious correlations since they have smaller supports. For $u_b$ and $u_0$, the sensitivity is more difficult to predict {\color{black}since the filtering problem is in a difficult regime for these two variables as explained in Section 4a in addition to highly correlated training data.}

  
The analysis and truth for the experiment ME2 {\color{black}$\rho=6$} with $K=10$ are compared in a series of plots in Figures \ref{comparison1} and \ref{comparison2}. A snapshot of the meridional profile of the zonal wind $u$, potential temperature $\theta$ and total heating $H$ are plotted in Figure \ref{comparison1}, with superimposed velocity field components $v$ and $w$. Although the analyzed velocity field $(v,w)$ is not well recovered, the analysis fields $u$ and $\theta$ appear to be closer to the true state. We should note here that the data assimilation experiments are performed in the presence of model error and among these variables, only $\theta$ is observed indirectly through (\ref{SW5}).  The total heating $H$, calculated from the non-assimilated heating rate modes $H_c$, $H_d$ and $H_s$, is expected to be harder to recover than the fully assimilated fields $\theta$ and $u$. Figure \ref{comparison2} contains the Hovm\"oller plots of the indirectly observed free tropospheric moisture $q$ and non-assimilated cloud area fraction $\sigma_d$. The analyzed $q$ compares well with the true state but the filter estimate of the field $\sigma_d$ contains errors relative to the true field. 

\section{Summary and conclusions}\label{SumCon}

In this paper, we demonstrated the efficacy of the localization maps introduced in \citep{de2017data} in a series of OSSEs realized with the monsoon-Hadley multicloud model, an idealized model with roughly three thousand six hundred model coordinates for the synoptic scale Hadley circulation and monsoonal flow. The model features a stochastic parameterization for clouds to represent the subgrid-scale processes of convection and precipitation and a bulk boundary layer dynamical model. We implemented the localization maps in a serial EnKF to assimilate satellite-like nonlinear indirect observations using an idealized radiative transfer model. We took vertically integrated brightness temperature measurements on 6 different channels over a sparse observational network (the total number of observations are close to 400). 

From the perfect model configuration, we learn that the data-driven localization map with small ensemble sizes of order $O(10)$ produced analysis estimates closer to those obtained from EnKF with ensemble sizes of order $O(1000)$ compared to the other methods in our numerical experiments, provided that the training data are close to stationary.  Of all 14 analyzed fields, the filter has most difficulty recovering the free troposphere barotropic zonal wind $u_0$ and the ABL zonal wind $u_b$ since {\color{black} these two slow variables that are not directly observed are in a difficult filtering regime, as explained in Section~4a. As a consequence, the training data of these two fields are highly correlated compared to the other fields. This means that for these two variables, the assumption for the training strategy, that is, stationarity on the correlation distribution, is violated. Nevertheless, when the ensemble size is extremely small, the numerical results suggest that the vector localization map, which uses information of adjacent spatial correlations to improve the correlation estimates, is more robust relative to the scalar localization map that is analogous to the usual Schur product-based localization function. In fact, our numerical results showed consistent improvement over the usual Gaspari-Cohn localization especially when small ensemble sizes are used.}

We also checked the proposed localization mapping in the presence of model error arising from misspecification of the convective time scales that impacts on the stochastic dynamics of the cloud area fractions, and in turn affects the large-scale through the convective closure of the model. In this scenario, we found that the filter performances using the localization maps obtained from imperfect model training data are almost identical to those using the localization maps obtained from perfect model training data. In some variables ($u_b, u_0, u_1, v_1, u_2$ and $\theta_1$), where most of these are not observed, we found that the localization mapping outperforms their own imperfect model training data assimilation filter skill. {\color{black} This result is possibly due to an improved effective ensemble size with the proposed localization mapping, which is analogous to the finding by \cite{oke2007} in a simpler context.}

Closer inspection reveals that the proposed least square fitting in \eqref{minprobdiscrete} is an ill-conditioned problem with condition numbers as large as $10^7$. This suggests that the quality of the data set is important for accurate estimation of the maps. {\color{black}While this is not a desirable feature, we found that this issue (which we encountered in the case of $K=30$) can be overcome by using a longer training data set, which offsets the larger Monte-Carlo error variance in correlation estimates between observation and model variables of large distances. Several strategies were proposed to overcome this issue when longer data set is not available. We numerically verified one of these strategies, namely by using the resulting maps optimized for smaller ensemble sizes. 

{\color{black} The numerical results from the scalar map $\rho=0$ is better compared to the standard GC localization. In particular, the filtering with scalar maps does not blow up in the case of small ensemble sizes where GC does. The major difference between these two localization techniques is that the scalar maps have non-uniform bandwidths while the GC localization uses a uniform bandwidth function. From a practical standpoint, these results are encouraging since specifying non-uniform bandwidths for the GC localization function is not trivial. On the other hand, the scalar maps are trained by setting one parameter $\rho=0$. Furthermore, the cost of training the scalar map is less than that of the vector map and the scalar maps are less sensitive to the training data compared to the vector maps (at least we didn't encounter the sensitivity issue as in the case of $\rho=6$ in our numerical experiments). For the vector map, besides the sensitivity issue, an adequate choice of parameter $\rho$ is needed to see the improvement as shown in our numerical example.}

From the encouraging results in this paper, this data-driven localization mapping is scalable for high dimensional applications, replacing the usual distance-based parametric-type localization function which is designed for spatial correlations that are local. The non-parametric nature of this approach allows the data to flexibly determine the appropriate non-trivial shape of the localization maps, including various non-local correlation dependence that is usually ignored with the standard localization. One potential issue is the availability of the high-quality training data set since generating training data without any localization (as done in this paper) is not possible for Atmospheric Global Circulation Models at this point. However, one can train the localization maps using empirical correlations obtained from large ensemble member data assimilation simulations with a very broad localization range such those demonstrated in \citep{miyoshi2014}. {\color{black}Another issue is the availability of the high-quality training data set in the presence of more severe modeling error, beyond parameter misspecification considered in this paper. In this situation, one may need more advanced model error estimation techniques \citep{harlim:2017} to generate reliable training data set. Another challenge in the operational setting is that the atmospheric dynamics are intermittent and seasonal. In addition, various types of observations are usually assimilated. It remains interesting to see whether we can use the idea in this paper to train the localization maps for the observations which have nontrivial non-local correlation structures and whether substantial improvement can be attained to offset the cost in the training procedure. 
}

%
\acknowledgments
The authors would like to thank Dr. Peter Houtekamer for his careful reading the manuscript and insightful comments. The research of J.H. is partially supported by the ONR Grant N00014-16-1-2888 and the NSF grants DMS-1317919, DMS-1619661. 

%






%
%
%

%
\newpage
\begin{table}[t]
\begin{center}
\begin{scriptsize}
\caption{Cloud transition probability rates $r_{kl}$ and timescales $\tau_{kl}$ in the stochastic multicloud parameterization. Here $C$, $C_l$ and $D$ are measures of the environment CAPE, low level CAPE, and midtroposphere dryness, respectively. The timescales' mean and standard deviation (SD) Bayes estimates are obtained from \citet{de2016stochasticity}.
 }\label{table0}
\begin{tabular}[t]{  l  l  c  } 
\toprule
\textbf{Cloud transition} & \textbf{Probability rate}  &\textbf{Timescale Bayes mean (SD) in hrs} \\
\midrule
Formation of congestus &   $r_{01} =\Gamma(C_l)\Gamma(D) / \tau_{01} $ & $31.789$ (4.795)   \\ 
Decay of congestus &  $r_{10} = \Gamma(D)/ \tau_{10}  $ &1.761 (0.224)\\
Conversion of congestus to deep \qquad \qquad\qquad  &  $ r_{12} =  \Gamma(C)\big( 1 - \Gamma(D)\big) / \tau_{12} $& 0.238 (0.001)\\
Formation of deep &  $r_{02} = \Gamma(C)\big( 1- \Gamma(D) \big)/ \tau_{02} $&11.821 (0.211) \\
Conversion of deep to stratiform &   $r_{23} = 1/\tau_{23} $& 0.2570 (0.0001)\\
Decay of deep &  $r_{20} =  \big( 1 - \Gamma(C) \big) / \tau_{20}$& 9.551 (13.146)  \\
Decay of stratiform &   $r_{30 } = 1/ \tau_{30} $& 1.021 (0.002)\\  
\midrule
\multicolumn{3} {c} {\textit{Arrhenius function} \hfill  $\Gamma (x) = \left\{ 1 - e^{-x}\,\,\mathrm{if}\,\, x>0,\,\, 0 \,\,\mathrm{otherwise} \right\}$} \\
\bottomrule
\end{tabular}
\end{scriptsize}
\end{center}
\end{table}

\begin{table}[t]
\begin{center}
\begin{scriptsize}
\caption{{\color{black}Description of the experiments in Section 4.}}\label{table1}
\begin{tabular}[t]{  l | l   } 
\toprule
\textbf{Map} & \textbf{\qquad \qquad \qquad \quad  Description}   \\ 
\midrule
\multicolumn{2} {c} {\qquad   Perfect model experiments (truth: regime A, forecast: regime A)} \\
\midrule
PM & {\color{black}Map} trained on TD\_PM's correlations     \\
\midrule
\multicolumn{2} {c} {\qquad    Model error experiments (truth: regime A, forecast: regime B)} \\
\midrule
ME1 & {\color{black}Map} trained on TD\_PM's correlations   \\
ME2 & {\color{black}Map} trained on TD\_ME's correlations  \\
\midrule
\multicolumn{2} {c} {\qquad    Training data } \\
\midrule
TD\_PM & Perfect model experiment with 1000 members  \\
TD\_ME & Model error experiment with 1000 members  \\
\bottomrule
\end{tabular}
\end{scriptsize}
\end{center}
\end{table}


%
\begin{figure}[t]
\begin{center}
\includegraphics[width=16cm]{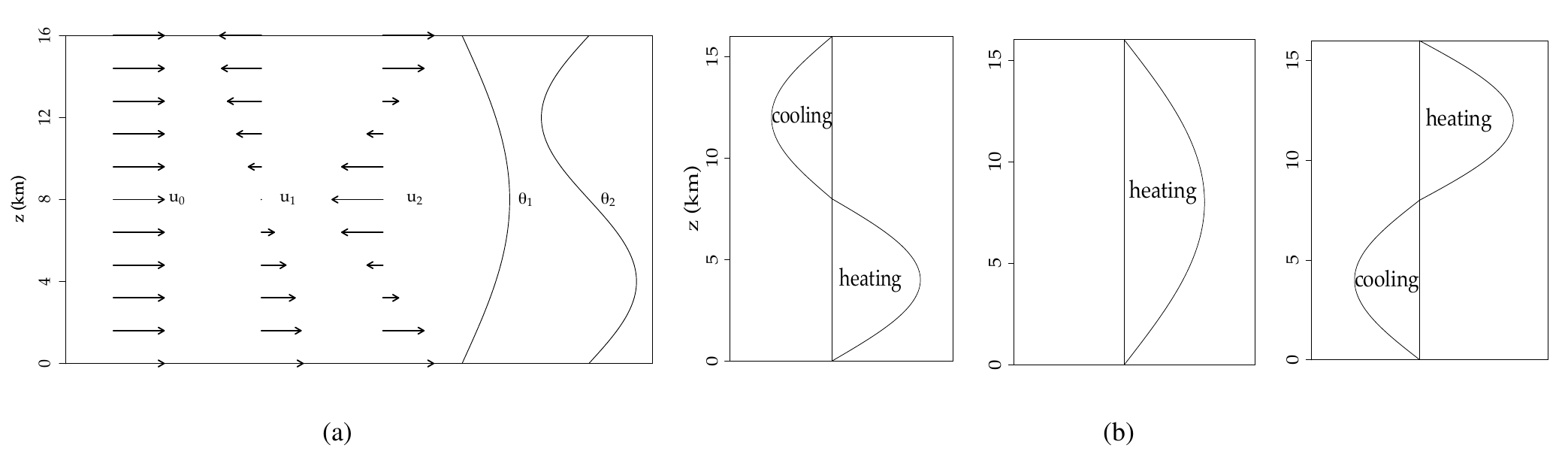}
\caption{{\bf (a)} Vertical profiles of the leading modes of horizontal velocity $\boldsymbol{u}$ and potential temperature $\theta$. (\textit{First three profiles}) Barotropic mode $\boldsymbol{u}_0$ and first two baroclinic modes of  velocity $\boldsymbol{u}_1$ and $\boldsymbol{u}_2$, respectively.  (\textit{Last two profiles}) First two baroclinic modes of temperature $\theta_1$ and $\theta_2$.  {\bf (b)} Baroclinic profile of the heating (and cooling) rates associated with the three cloud types of the multicloud model. (\textit{Left}) Mode--2 congestus heating $H_c$. (\textit{Center}) Mode--1 deep heating $H_d$. (\textit{Right}) Mode--2 stratiform heating $H_s$. The heating curves intersect the vertical straight lines at zero heating points. }\label{fig_vertical_struc}
\end{center}
\end{figure}

\clearpage
\begin{figure}[t]
\begin{center}
\includegraphics[width=8cm]{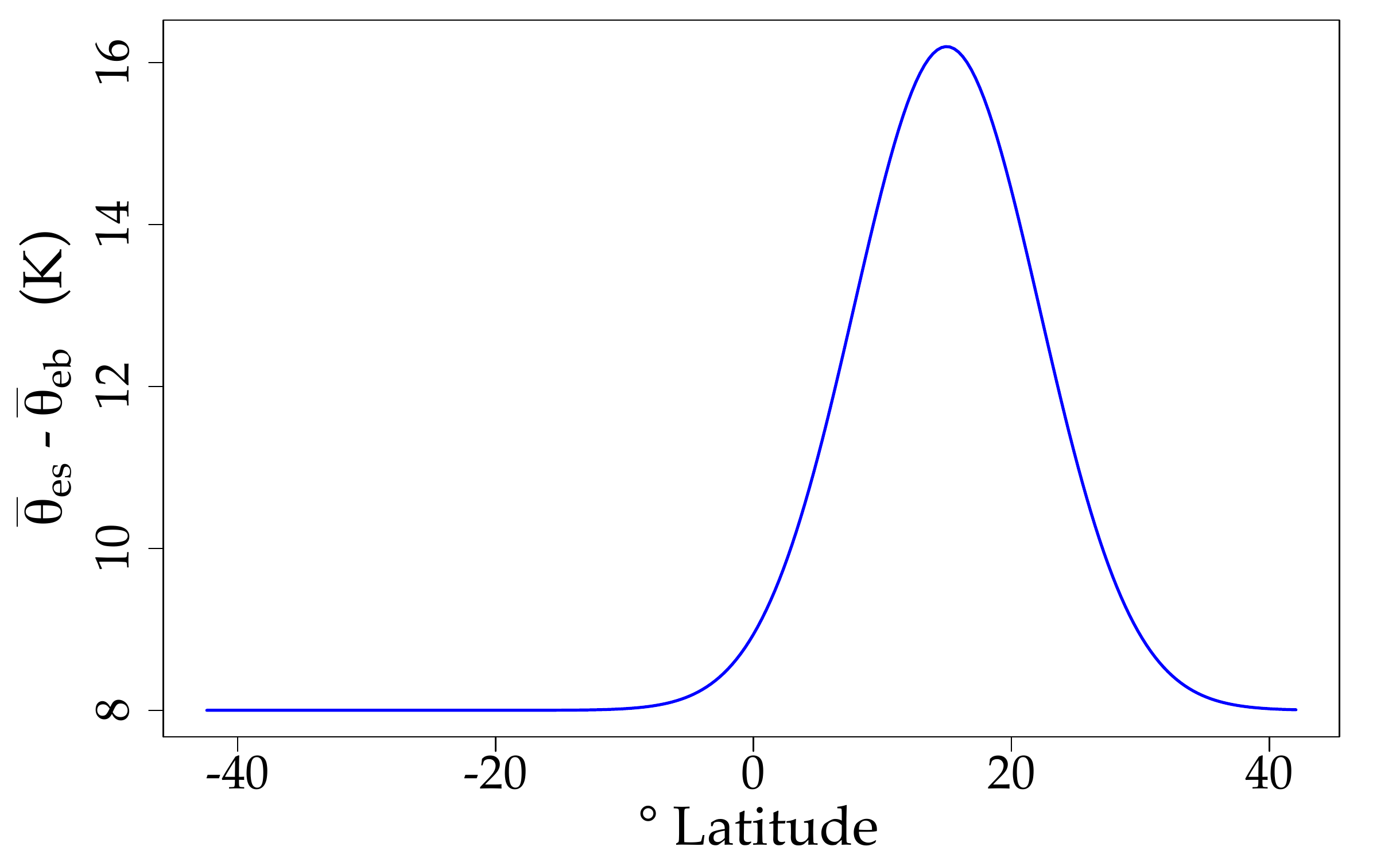}
\vspace{-10pt}
\caption{Imposed SST meridional profile. The surface temperature gradient at RCE follows a normal distribution centered at $15^\circ$ N with a standard deviation of $7.2^\circ$. $\overline{\theta}_{es}$  and $\overline{\theta}_{eb}$ are the surface and ABL equivalent potential temperatures at RCE, respectively.  } \label{temp_profile}
\end{center}
\end{figure}

\begin{figure}[t]
\begin{center}
\includegraphics[width=16cm]{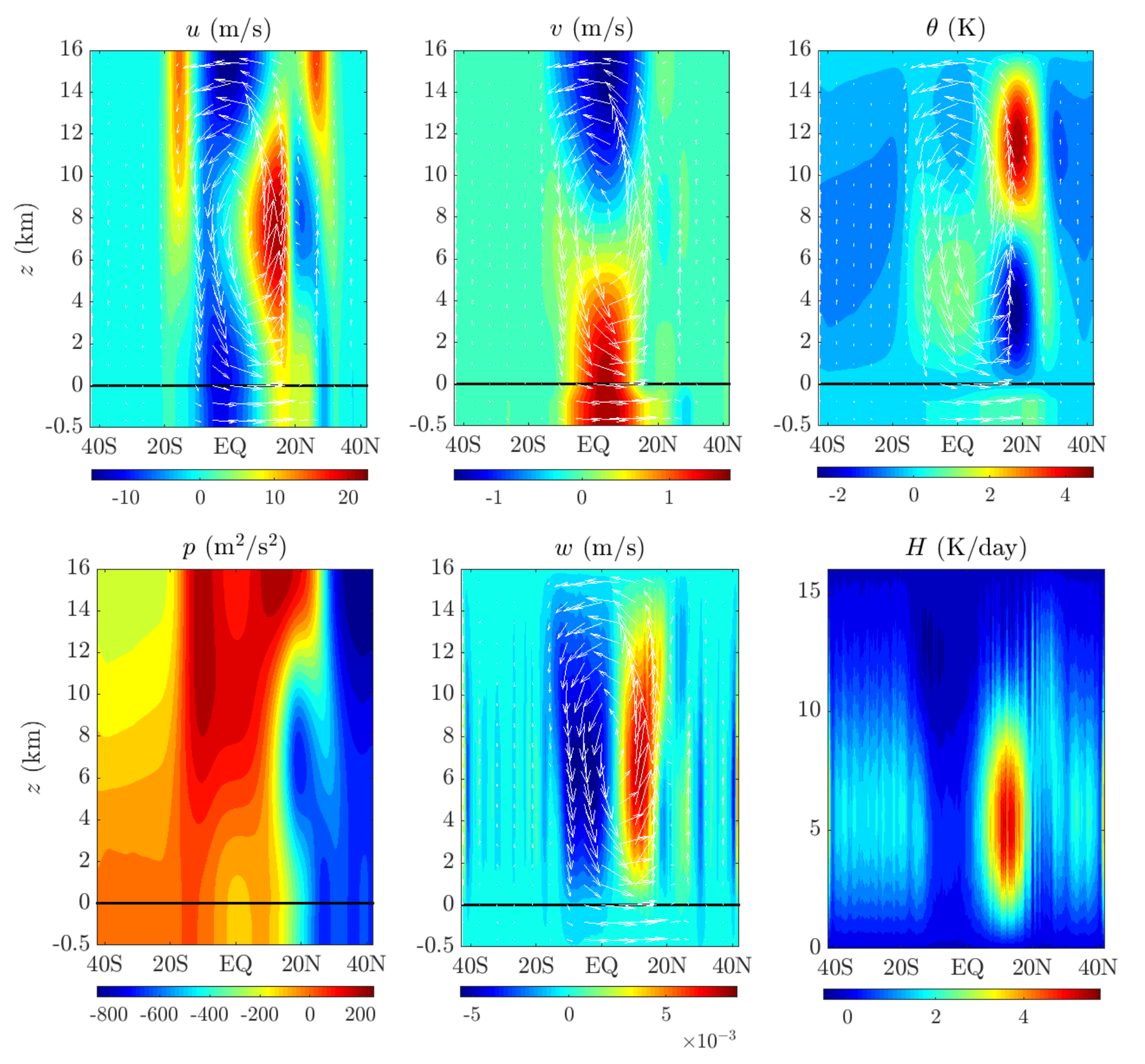} 
\caption{Mean meridional circulation averaged over the training interval for regime A. The top of the ABL (solid black line) is located at height 0 km. The contours represent the indicated fields, and the arrows are the velocity vector field ($v$,$w$). Clockwise: Zonal and meridional winds, potential temperature, total heating, vertical velocity and pressure. }\label{meancirc}
\end{center}
\end{figure}

\begin{figure}[t]
\begin{center}
\includegraphics[width=16cm]{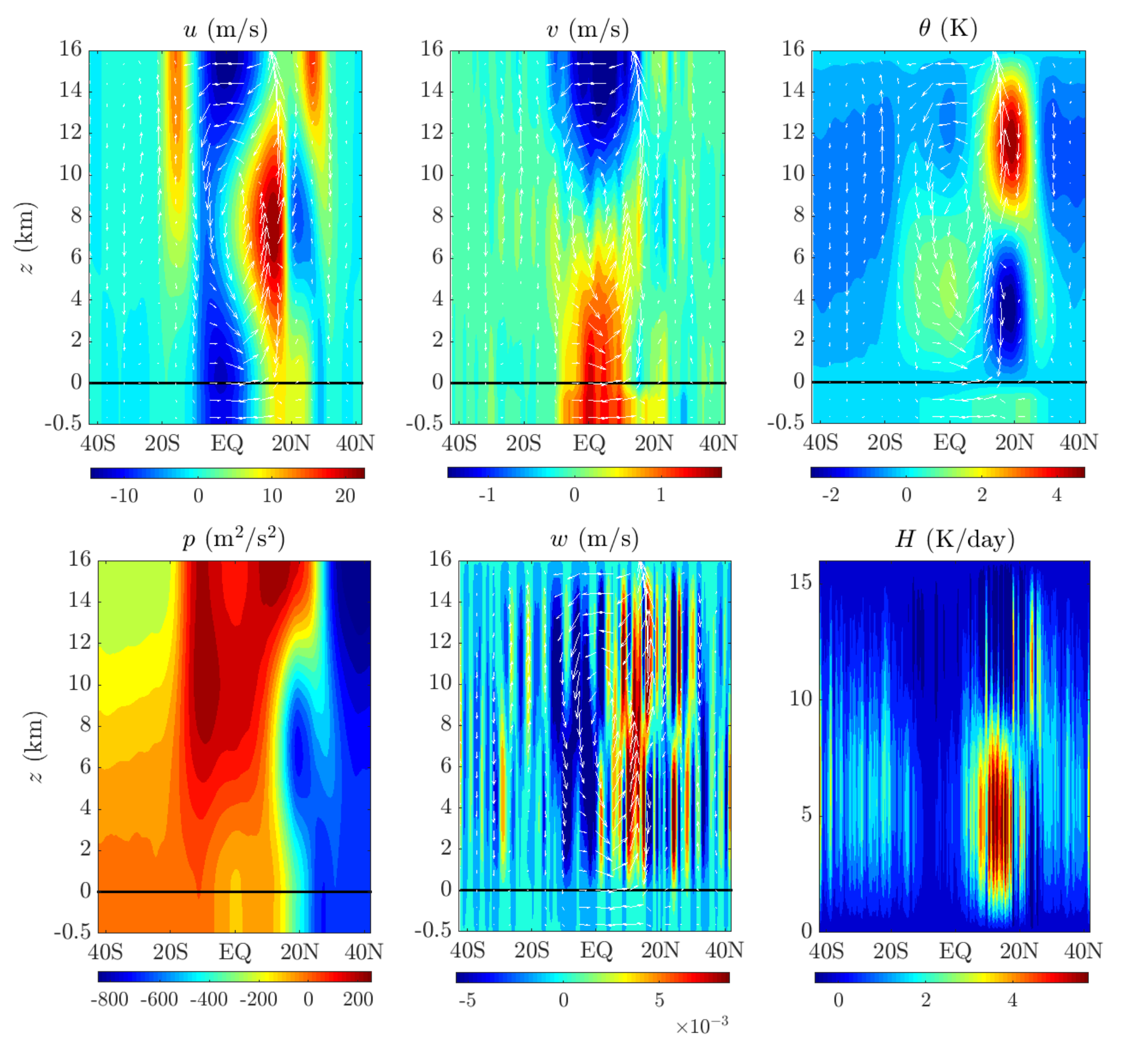} 
\caption{Same as Fig. \ref{meancirc} but for regime B. }\label{meancirc2}
\end{center}
\end{figure}

\begin{figure}[t]
\begin{center}
\includegraphics[width=16.cm]{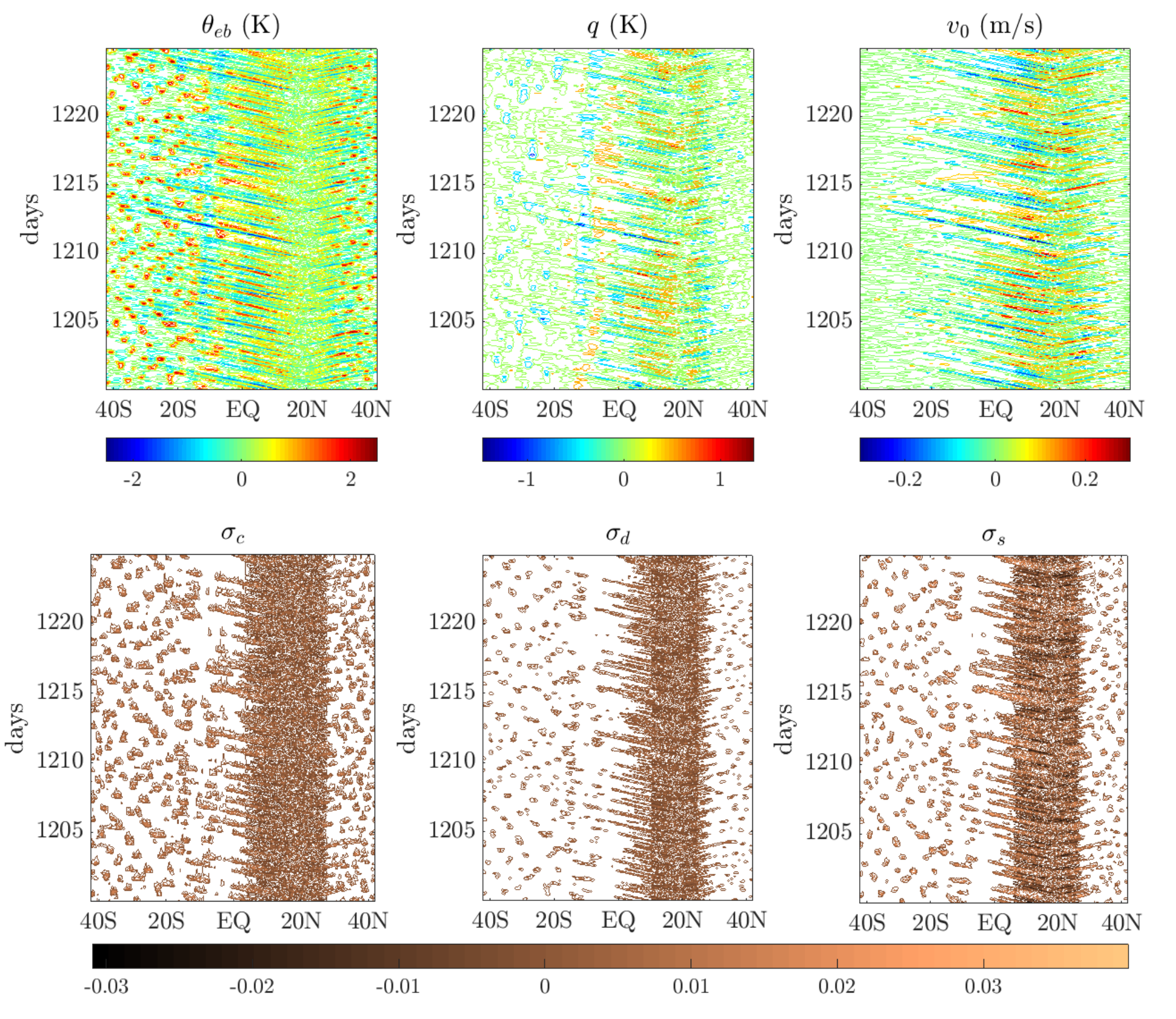} 
\caption{25-day Hovm\"oller plots for regime A. Clockwise: ABL equivalent potential temperature, free tropospheric moisture, meridional barotropic wind; Stratiform, deep and congestus cloud area fractions. } \label{waves}
\end{center}
\end{figure}

\begin{figure}[t]
\begin{center}
\includegraphics[width=16.cm]{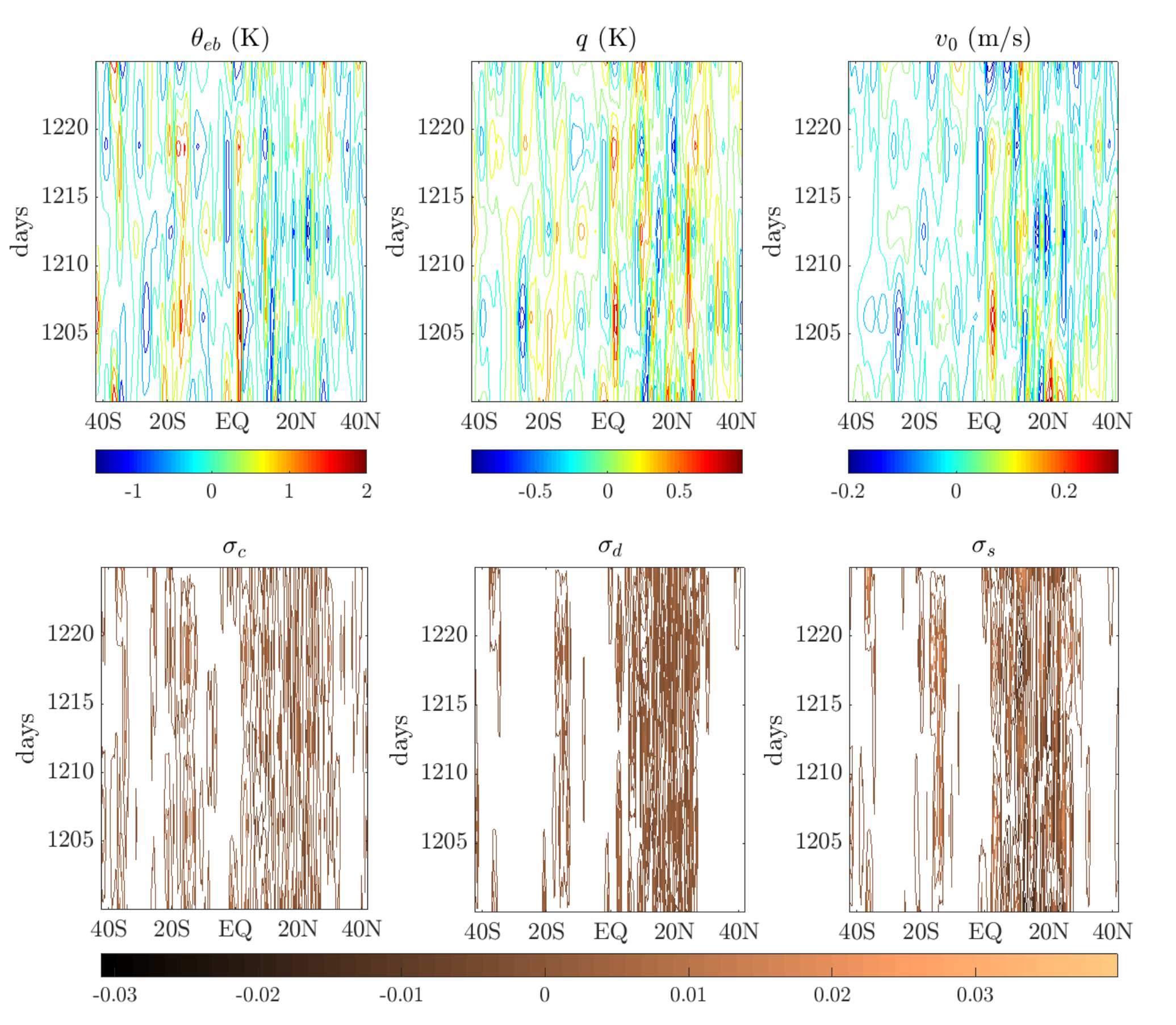}  
\caption{Same as Fig. \ref{waves} but for regime B. } \label{waves2}
\end{center}
\end{figure}

\begin{figure}[t]
\begin{center}
\includegraphics[width=12cm]{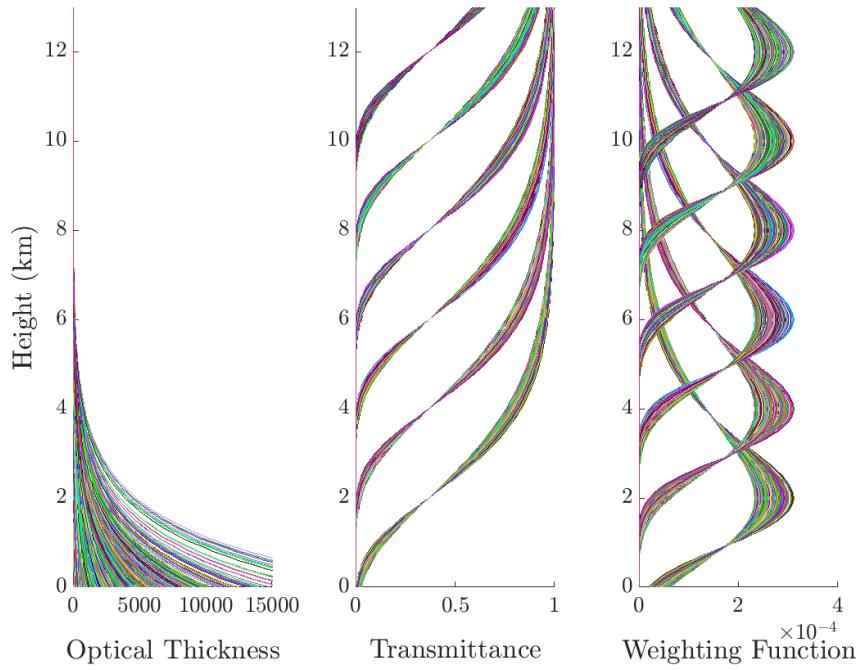}   
\caption{Optical thickness $\tau(q,\infty,z)$ (\textit{left}), transmittance $T_{\lambda}(q,\infty,z)$ (\textit{middle}) and weighting function $K_{\lambda}(q,\infty,z)$ (\textit{right}) {\color{black}as a function of height $z$} for the 6 channels $\lambda_1$, $\dots$, $\lambda_6$, shown here (in an array of colors; quantities are nondimensional) for different {\color{black}climatological moisture values $q$}.  }\label{SixChannels}
      \end{center}
\end{figure}

\clearpage
\begin{figure}[t]
\begin{center}
\includegraphics[width=10cm,height=6cm]{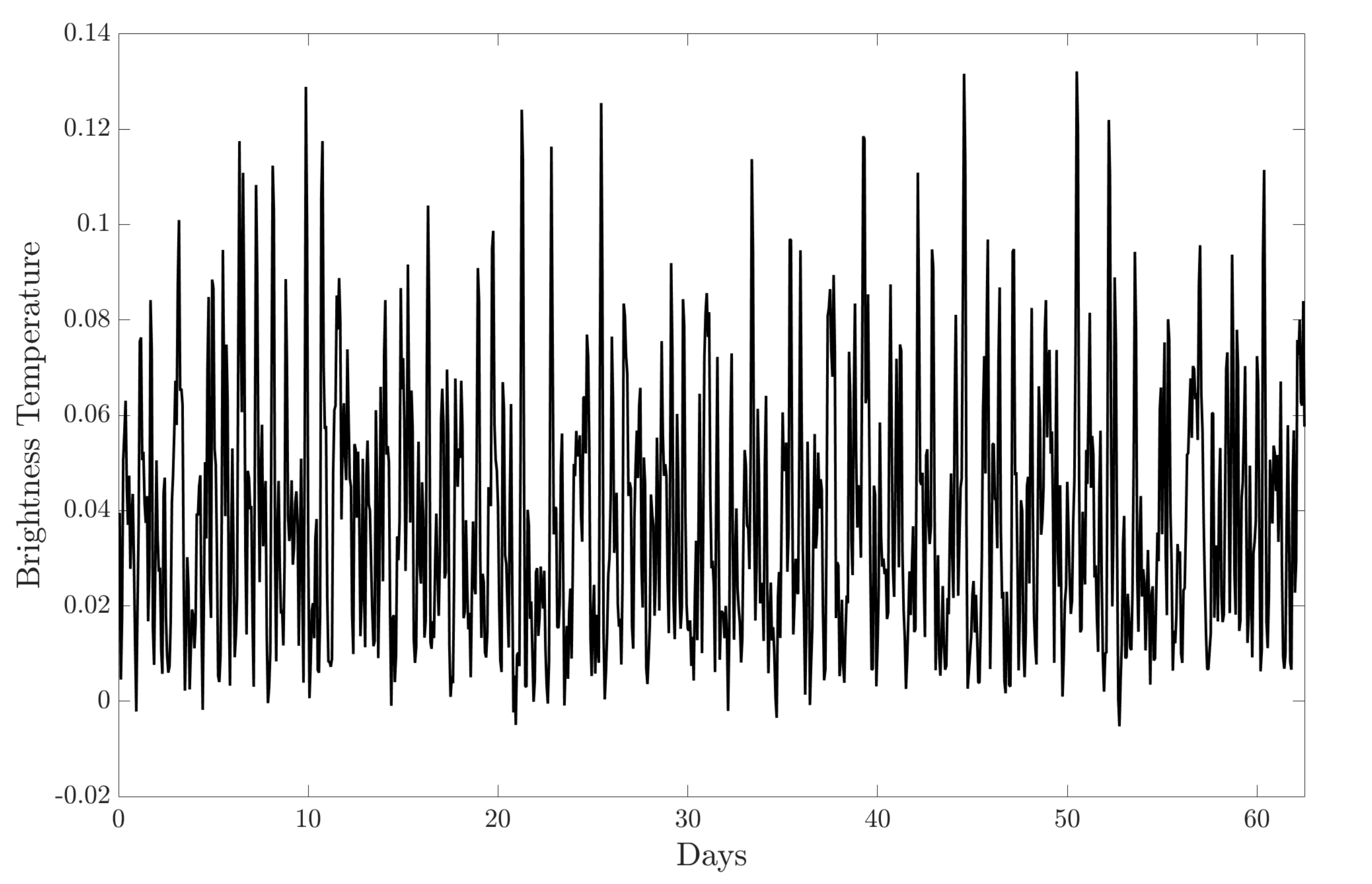}   
\caption{Nondimensional brightness temperature $T_{b,{\lambda_1}}$ associated with the first channel, calculated  from climatology over a period of 65 days.  }\label{BT}
      \end{center}
\end{figure}

\clearpage
\begin{figure}[t]
\begin{center}
\includegraphics[width=1\textwidth]{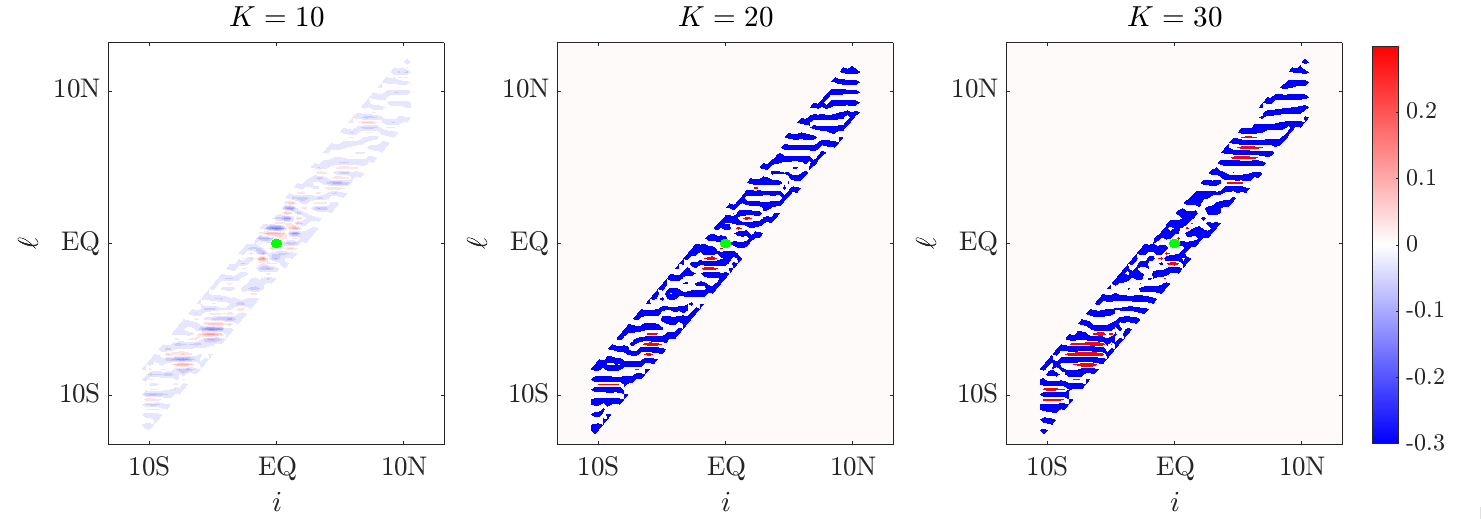} 
\caption{Contour plot of the vector localization map for the correlations between the model coordinates $i$ of the field $u_b$ and a Channel 1 brightness temperature observation $j$ located at the green point (near the Equator). The contour value at the coordinates $(i,\ell)$ corresponds to the component $\ell$ of the vector map $\mathcal{L}_{ij}$. For a fixed $i$ and $j$, the vector components of $\mathcal{L}_{ij}$ (along the $y$-axis) are zero outside the radius $\rho=6$ of the observation location.} \label{lmap_contour}
\end{center}
\end{figure}

\clearpage
\begin{figure}[t]
\begin{center}
\includegraphics[width=1\textwidth]{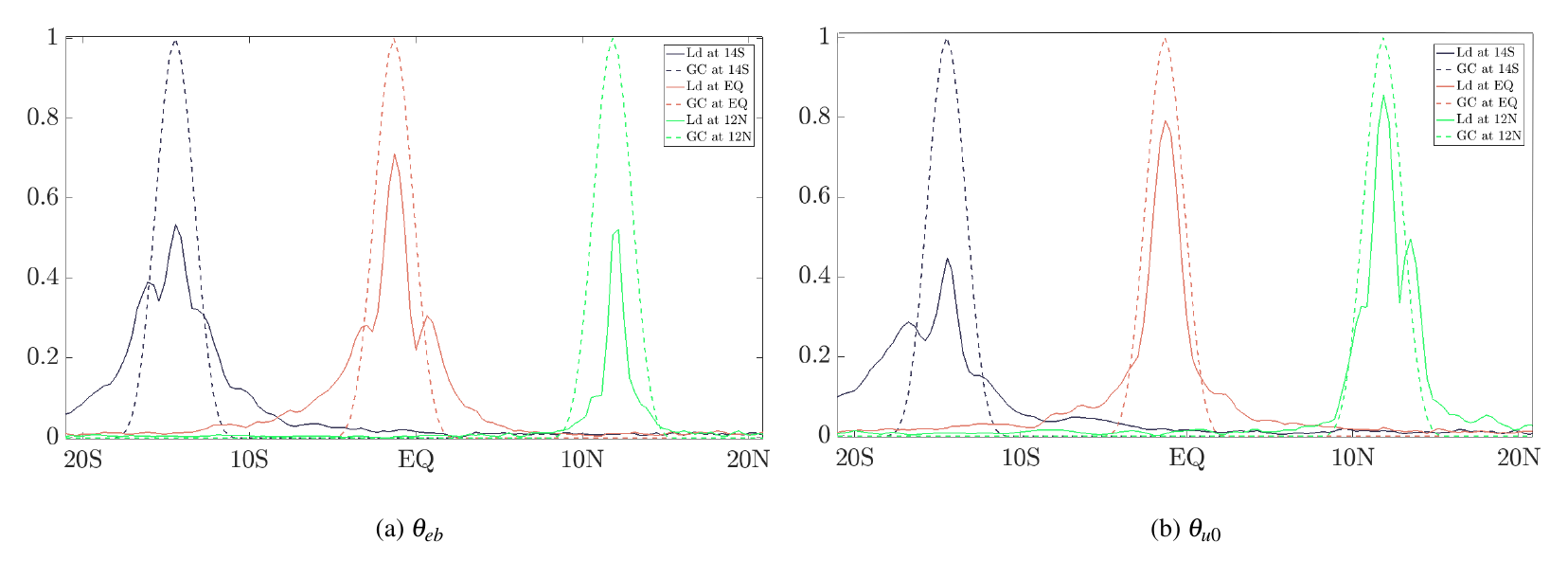} 
\caption{The GC function and localization map for the correlation between the observed brightness temperature of Channel 1 observed at three different locations (roughly $14^\circ $S, EQ and $12^\circ $N)  and the model variables (a) $\theta_{eb}$ and (b) $u_0$. } \label{GCvsLd}
\end{center}
\end{figure}

\clearpage
\begin{figure}[t]
\begin{center}
\includegraphics[width=0.6\textwidth]{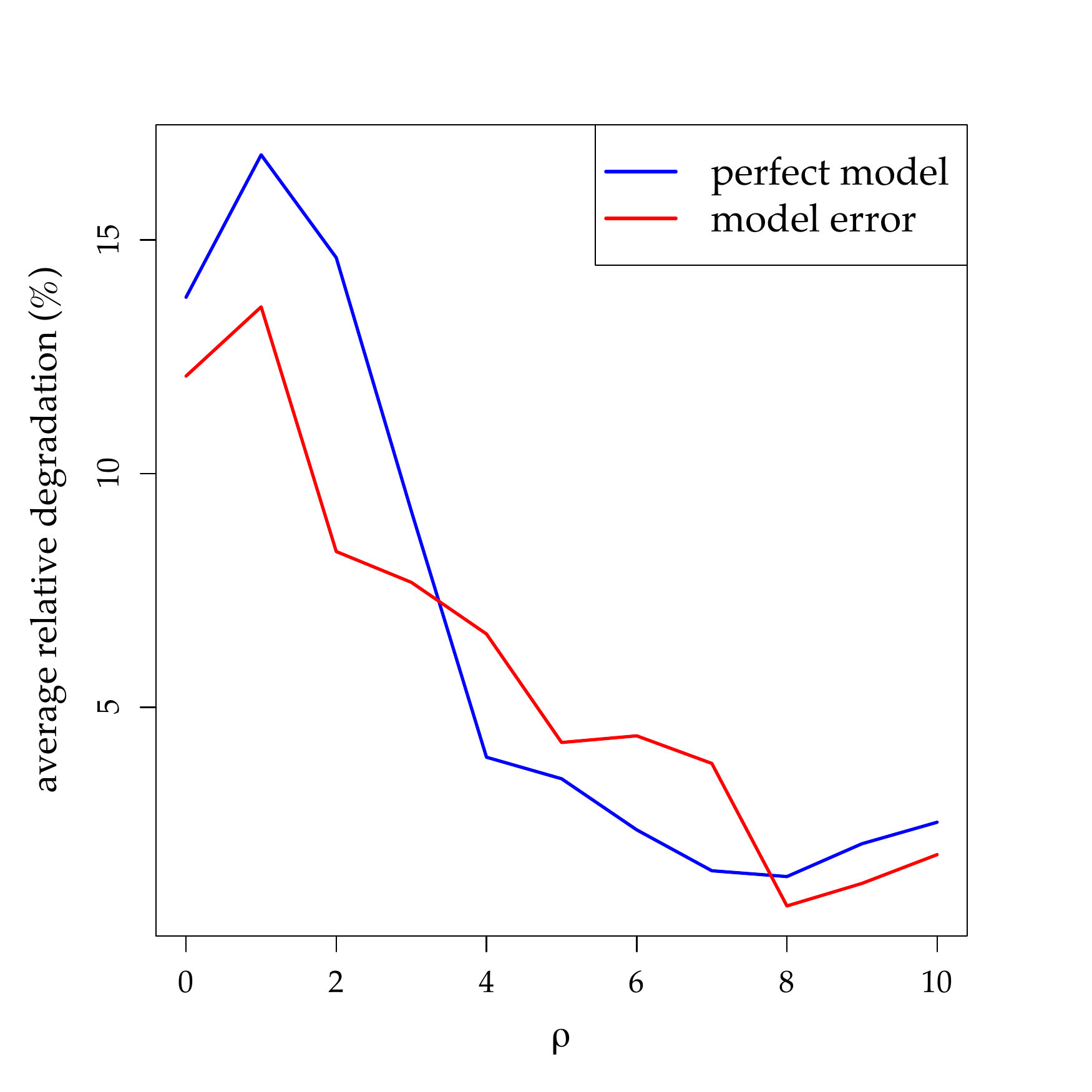}
\caption{Average relative degradation as a function of the parameter $\rho$ for the perfect model and model error experiments.} \label{degradation}
\end{center}
\end{figure}

\begin{figure}[t]
\begin{center}
\includegraphics[width=1\textwidth]{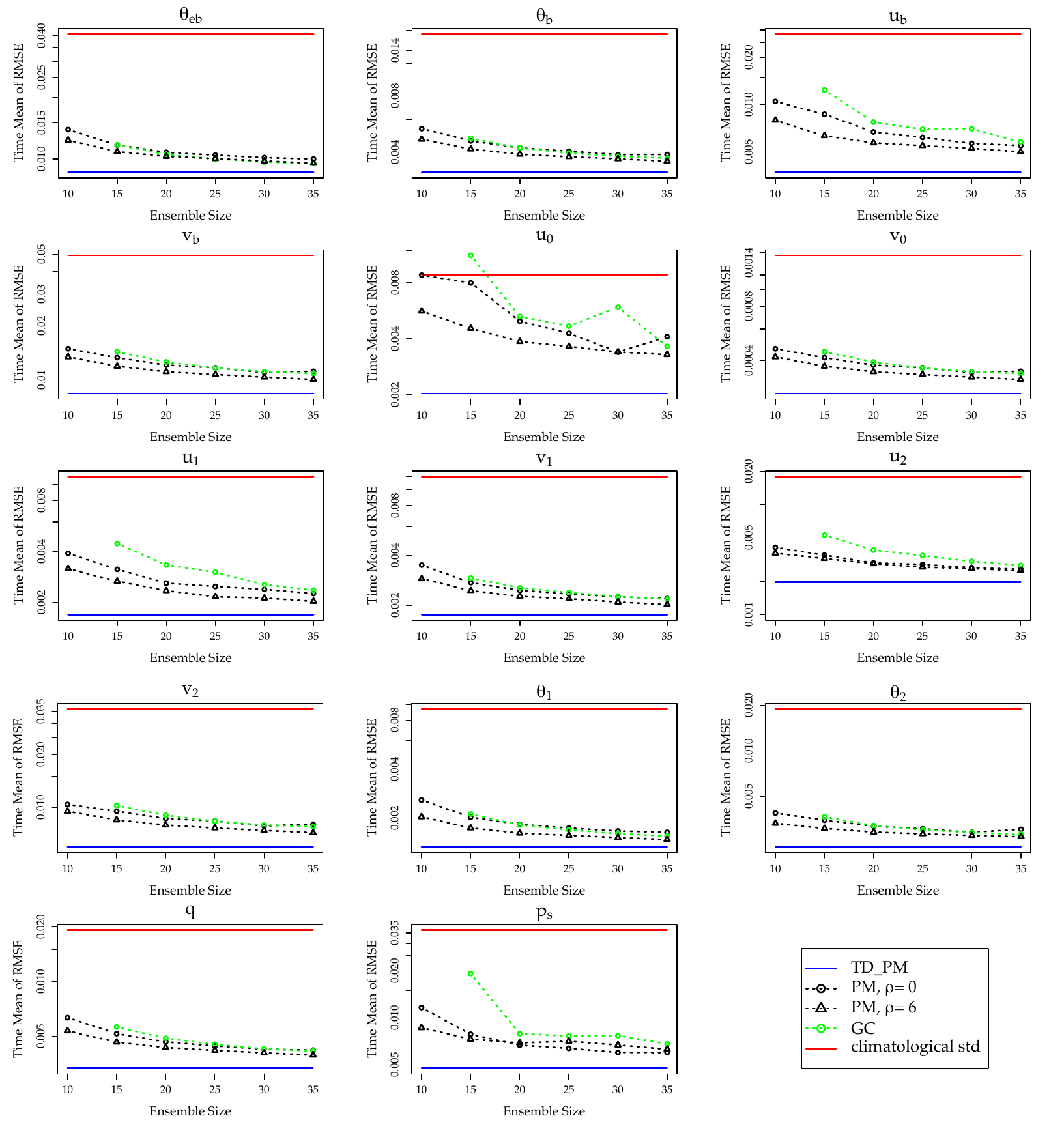}
\caption{\textit{Perfect model experiments.}  
Log-linear plot of the time mean of analysis RMSE as a function of ensemble size is shown for the 14 analyzed fields, for the verification experiments using the localization maps {\color{black}PM with $\rho=0$ and $\rho=6$} (dashed black). The training experiment TD\_PM using 1000 members is reported as a baseline (solid blue).  } \label{perfectmodel}
\end{center}
\end{figure}

\begin{figure}[t]
\begin{center}
\includegraphics[width=0.98\textwidth]{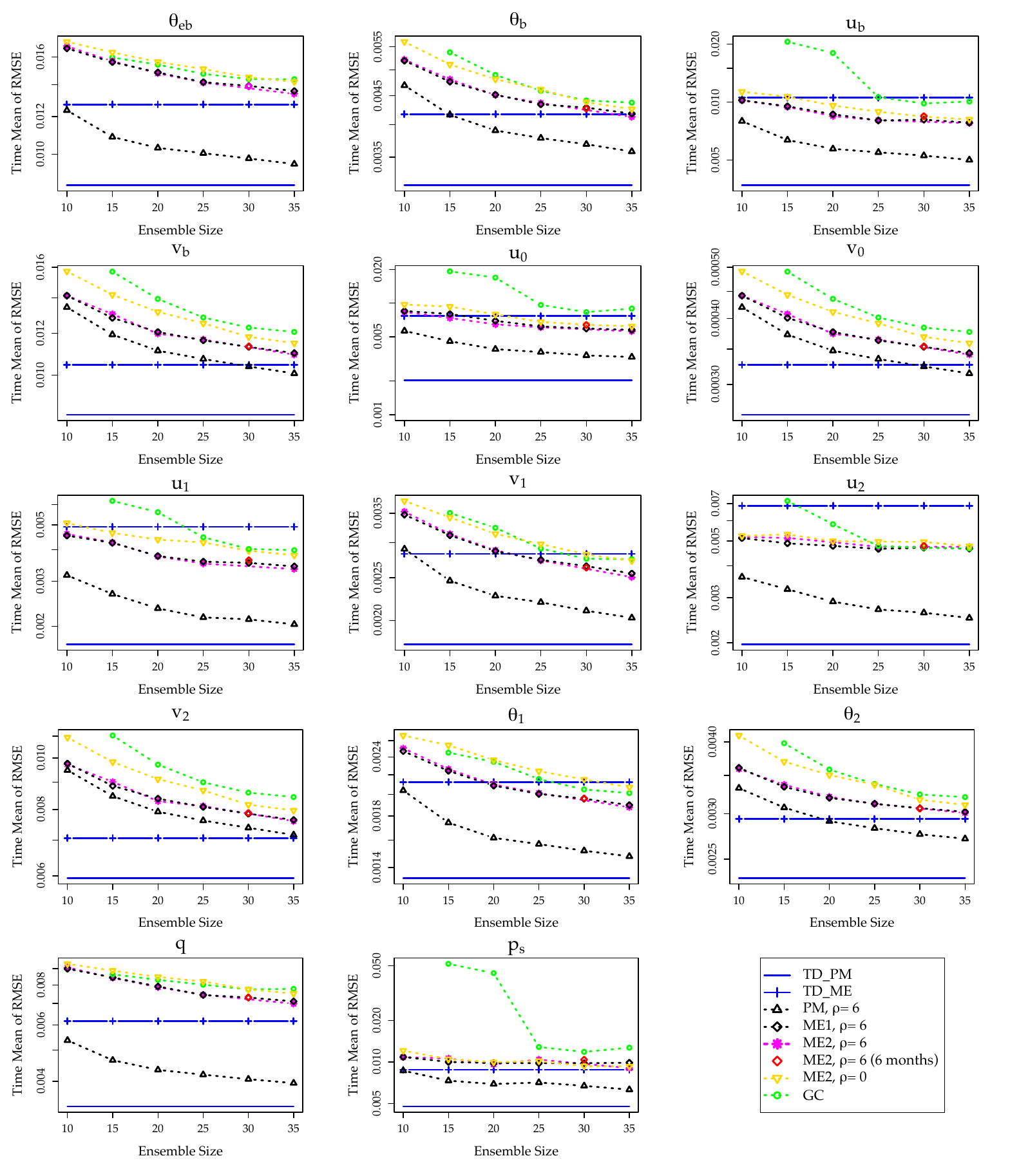}
\vspace{-0.6cm}
\caption{\textit{Model error experiments.}  
Log-linear plot of the time mean of analysis RMSE as a function of ensemble size is shown for the 14 analyzed fields, for the verification experiments using the localization maps ME1 and ME2. The training experiments TD\_ME and TD\_PM using 1000 members are reported as baselines (solid blue). } \label{modelerror}
\end{center}
\end{figure}

\begin{figure}[t]
\begin{center}
\includegraphics[width=.8\textwidth]{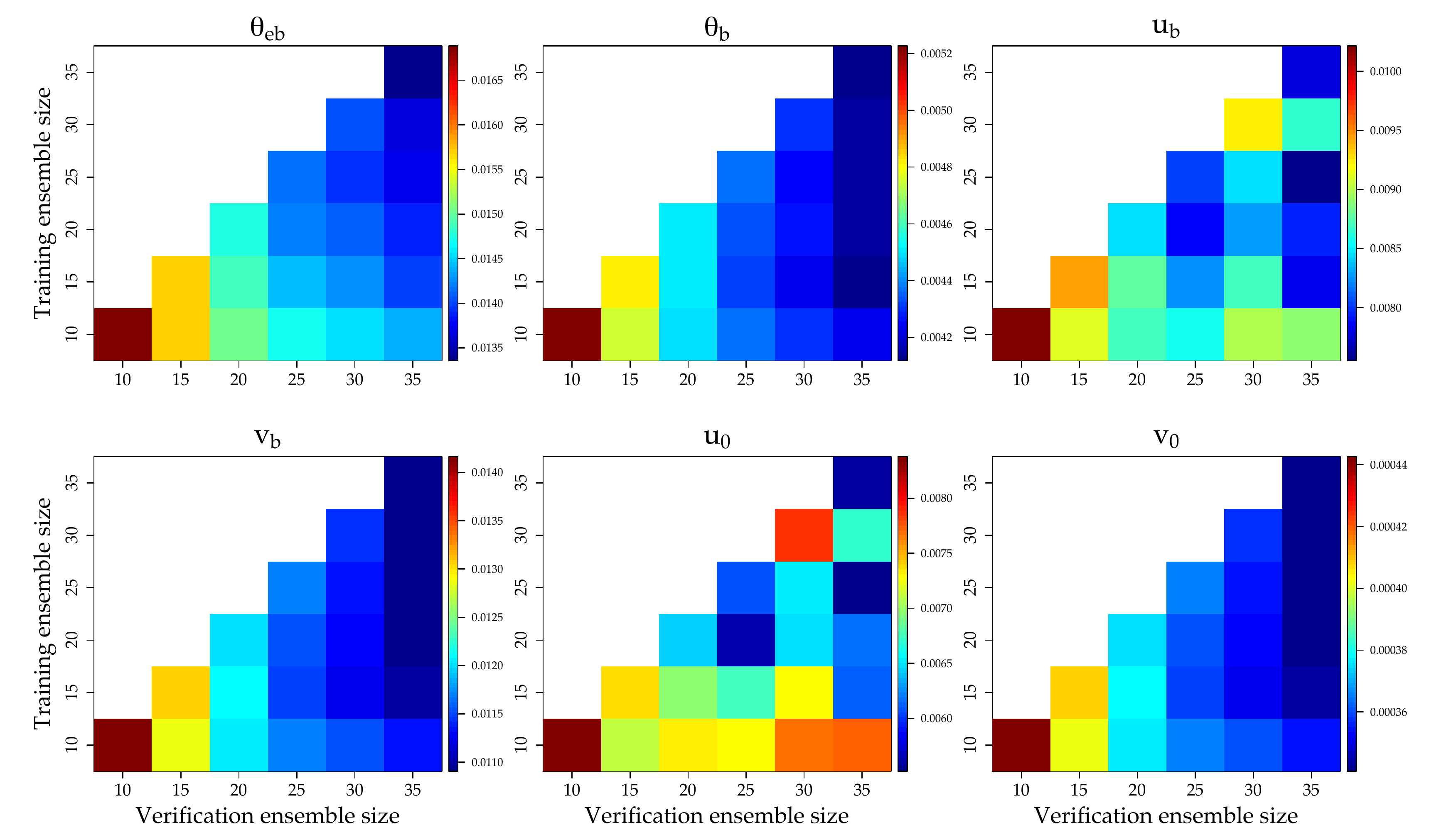}
\caption{\textit{Model error experiments}. Time mean of analysis RMSE as a function of training and verification ensemble sizes for experiment ME2 {\color{black}($\rho=6$)}. White pixels indicate filter divergence. } \label{swap}
\end{center}
\end{figure}

\begin{figure}[t]
\begin{center}
\includegraphics[width=.8\textwidth]{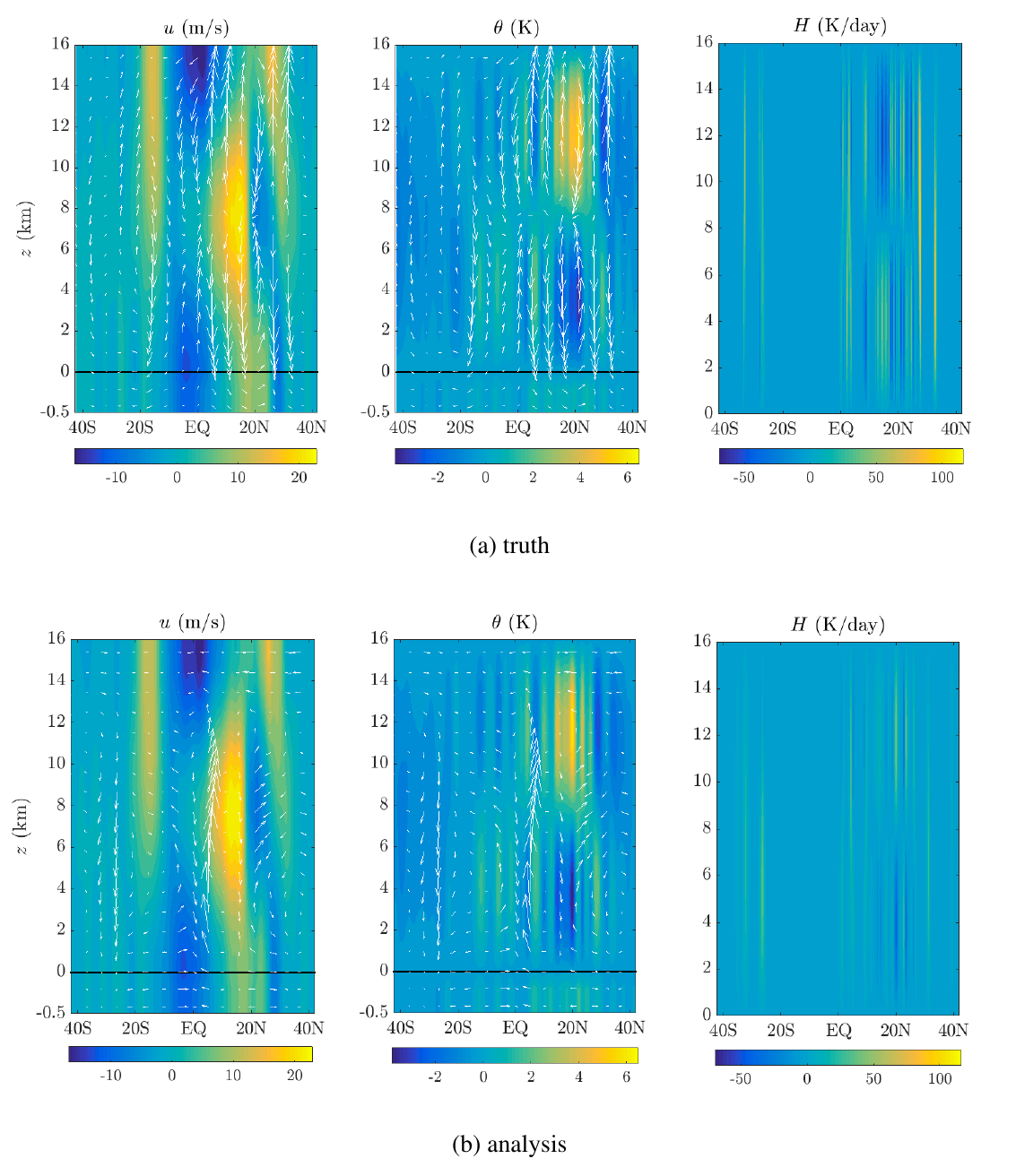}
\caption{Snapshot of the meridional circulation of (a) the truth and (b) the analysis for the verification experiment ME2 ({\color{black}$\rho=6$ and} $K=10$). The snapshot is taken at an arbitrary assimilation cycle. The contours represent the indicated fields, and the arrows are the velocity vector field $(v,w)$.  } \label{comparison1}
\end{center}
\end{figure}

\begin{figure}[t]
\begin{center}
\includegraphics[width=.8\textwidth]{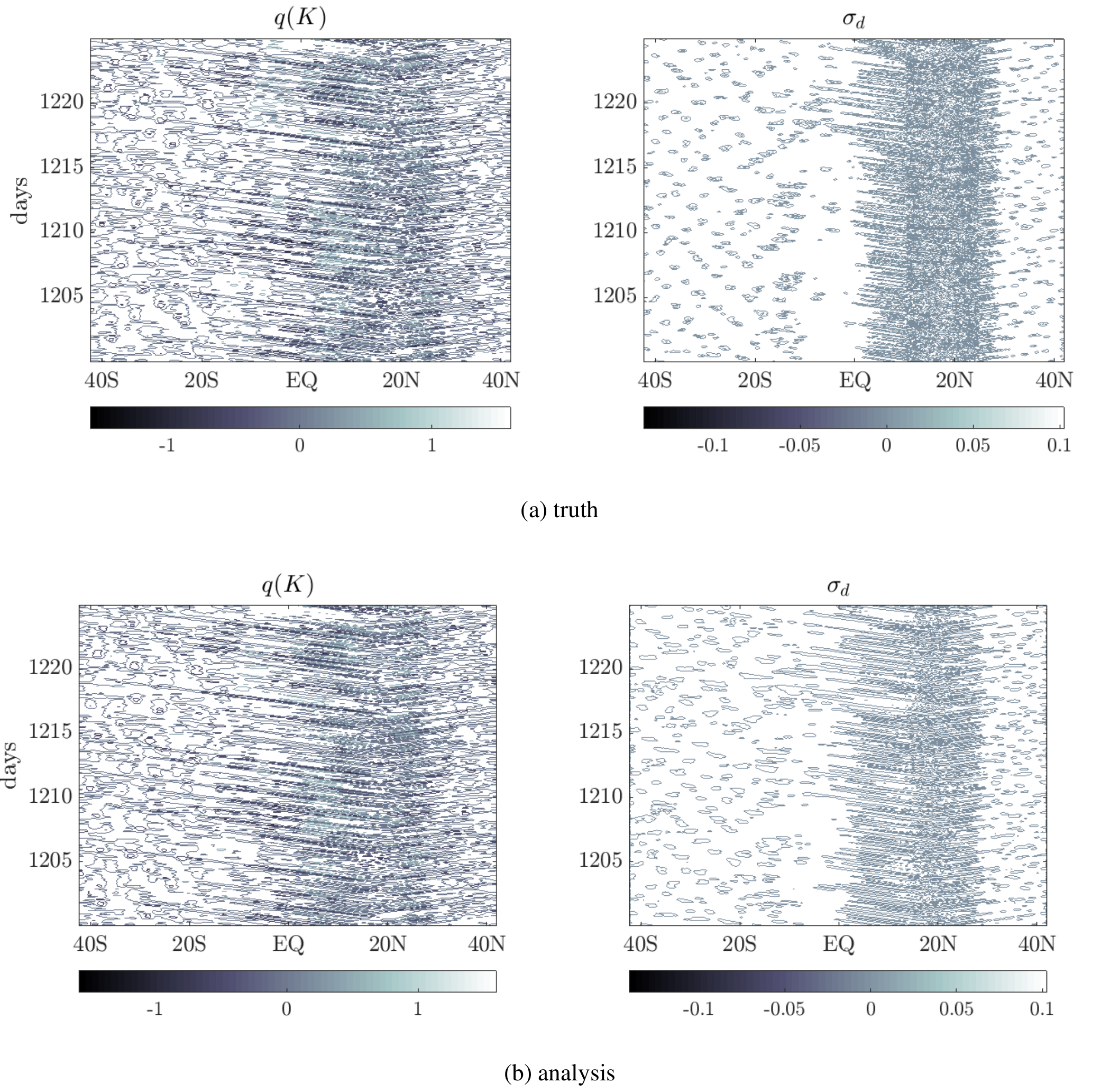}
\caption{25-day Hovm\"oller plots of (a) the truth and (b) the analysis for the verification experiment ME2 ({\color{black}$\rho=6$ and} $K=10$). (\textit{Left column}.) Free tropospheric moisture. (\textit{Right column}.) Deep cloud area fraction.} \label{comparison2}
\end{center}
\end{figure}


\end{document}